\documentclass[aps,pra,floatfix,twocolumn,superscriptaddress]{revtex4}

\usepackage{graphicx}
\usepackage{graphics}
\usepackage{amssymb}
\usepackage{amsmath} 
\usepackage{epsfig}
\usepackage{latexsym}
\usepackage{color}
\usepackage{subfigure}
\usepackage{array}

\newcommand{\ket}[1]{|#1\rangle}
\newcommand{\bra}[1]{\langle #1|}
\newcommand{\braket}[2]{\langle #1|#2\rangle}

\begin{document}

\title{Fast Non-Adiabatic Two Qubit Gates for the Kane Quantum Computer}

\author{Charles D. Hill} 
\email{hillcd@physics.uq.edu.au}
\affiliation{Centre for Quantum Computer
Technology, and Department of Physics, The University of Queensland, St
Lucia, QLD 4072, Australia} 

\author{Hsi-Sheng Goan}
\email{goan@physics.uq.edu.au}
\affiliation{Centre for Quantum Computer Technology, University of
New South Wales, Sydney, NSW 2052, Australia}
\thanks{Mailing Address: Centre for Quantum Computer
Technology, C/- Department of Physics, The University of Queensland, St
Lucia, QLD 4072, Australia}


\begin{abstract}
In this paper we apply the canonical decomposition of two qubit
unitaries to find pulse schemes to control the proposed Kane quantum
computer. We explicitly find pulse sequences for the CNOT, swap,
square root of swap and controlled Z rotations. We analyze the speed
and fidelity of these gates, both of which compare favorably to
existing schemes. The pulse sequences presented in this paper are
theoretically faster, higher fidelity, and simpler. Any two qubit gate
may be easily found and implemented using similar pulse
sequences. Numerical simulation is used to verify the accuracy of each
pulse scheme.
\end{abstract}

\maketitle


\section{Introduction}

The advent of quantum algorithms \cite{Gro97, Sho97} that can
out-perform the best known classical algorithms has inspired many
different proposals for a practical quantum computer \cite{KLM01,
GC97, CFH97, CZ95, NPT98, IAB+99, Kan98}. One of the most promising proposals
was presented by Kane \cite{Kan98}. In this proposal a solid state
quantum computer based on the nuclear spins of $\mathrm{^{31}P}$ atoms was
suggested. Although initially difficult to fabricate, this scheme has
several advantages over rival schemes \cite{KLM01, GC97, CFH97, CZ95,
NPT98, IAB+99}. These include the comparatively long decoherence times
of the $\mathrm{^{31}P}$ nuclear and electron spins \cite{Hon54, GB58,
FG59, Feh59, HSE60, Fau69, CH72, WJS88}, the similarity to existing
$\mathrm{Si}$ fabrication technology, and the ability to scale.

There have been two main proposals for pulse sequences to implement a
CNOT gate on the Kane quantum computer. In the initial proposal
\cite{Kan98} an adiabatic CNOT gate was suggested. Since that time the
details of this gate have been investigated and optimized \cite{GM00,
Wel01, WCJ01, FWH03}. This adiabatic scheme takes a total time of
approximately $26 \mu s$ and has a systematic error of approximately
$5\times 10^{-5}$ \cite{Wel01}. As good as these results are,
non-adiabatic gates have the potential to be faster with higher fidelity
and allow advanced techniques such as composite rotations and modified
RF pulses \cite{CJ99, Tyc83}.

Wellard et al. \cite{WCJ01b} proposed a non-adiabatic pulse scheme
for the CNOT and swap gates. They present a CNOT gate that takes a
total time of approximately $80 \mu s$ with an error (as defined later
in Eq.\ (\ref{eqn:Error})) of approximately $4 \times
10^{-4}$. Although this gate is non-adiabatic it is slower than its
adiabatic counterpart. For the non-adiabatic swap gate a total time
was calculated of $192 \mu s$.

One of the most useful tools in considering two qubit unitary
interactions is the canonical decomposition \cite{KC01, HVC02,
KBG00}. This decomposition expresses any two qubit gate as a product
of single qubit rotations and a simple interaction content. The
interaction content can be expressed using just three parameters. In
the limit that single qubit rotations take negligible time (in
comparison to the speed of interaction), this decomposition can be
used to find optimal schemes \cite{HVC02, KBG00}, and of particular
inspiration to this paper is an almost optimal systematic method to
construct the CNOT gate \cite{BDD+02}.

It is not possible to apply those optimal schemes \cite{HVC02, KBG00}
directly to the Kane quantum computing architecture. They assume
single qubit gates take negligible time in comparison with two qubit
interactions, whereas on the Kane architecture, they do not. Secondly,
in the proposal for the Kane computer, adjacent nuclei are coupled via
the exchange and hyperfine interactions through the electrons, rather
than directly, and so we have a four `qubit' system (two electrons and two
nuclei) rather than a two qubit system. Although we cannot
apply \emph{optimal} schemes directly, in this paper we use the
canonical decomposition to simplify two qubit gate design.

Apart from being simple to design and understand, gates described in
this paper have many desirable features. Some features of these gates
include:

\begin{enumerate}
\item They are simpler, higher fidelity and faster than existing
proposals.
\item They do not require sophisticated pulse shapes, such as are
envisioned in the adiabatic scheme, to implement.
\item \emph{Any} two qubit gate can be implemented directly using
similar schemes. This allows us to implement gates directly rather
than as a series of CNOT gates and single qubit rotations.
\end{enumerate}

This paper is organized as follows. Sec.\ \ref{sec:Kane} gives an
overview of the Kane quantum computer architecture and single qubit
rotations. Sec.\ \ref{sec:Canonical} describes the canonical
decomposition as it applies to the Kane quantum computer. Section
\ref{sec:CNOT} describes pulse schemes for Control Z gates and CNOT
gates. Sec.\ \ref{sec:Swap} gives potential pulse schemes for swap
and square root of swap gates. Finally, the conclusion, Section
\ref{sec:Conclusion}, summarizes the findings of this paper.

\section{The Kane Quantum Computer}\label{sec:Kane}

\subsection{The Kane Architecture}

\begin{center}
\begin{figure}[h]
\includegraphics[width=8cm]{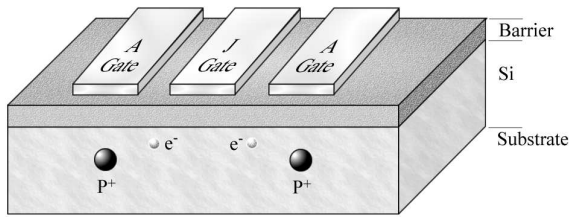}
\caption{The Kane Quantum Computer Architecture}\label{fig:Kane} 
\end{figure}
\end{center}

A schematic diagram of the Kane quantum computer architecture is shown
in Fig.\ \ref{fig:Kane}. The short description given here follows Goan
and Milburn \cite{GM00}. This architecture consists of
$\mathrm{^{31}P}$ atoms doped in a purified $\mathrm{^{28}Si}$ ($I=0$)
host. Each P atom has nuclear spin of $I=\frac{1}{2}$. Electrodes
placed directly above each P atom are referred to as A-Gates, and
those between atoms are referred to as J-Gates. An oxide barrier
separates the electrodes from the $\mathrm{P}$ doped $\mathrm{Si}$.

Each $\mathrm{P}$ atom has five valence electrons. As a first
approximation, four of these electrons form covalent bonds to
neighboring $\mathrm{Si}$ atoms, with the fifth forming a
hydrogen-like S-orbital around each $\mathrm{P^+}$ ion. This electron
is loosely bound to the $\mathrm{P}$ donor and has a Bohr radius of
$a_{B}^{*} \approx 3 \mathrm{nm}$, allowing an electron mediated
interaction between neighboring nuclei.

In this paper nuclear spin states will be represented by the states
$\ket{1}$ and $\ket{0}$. Electronic spin states will be represented by
$\ket{\uparrow}$ and $\ket{\downarrow}$. Where electronic states are
omitted, it is assumed that they are polarized in the
$\ket{\downarrow}$ state. $X$, $Y$ and $Z$ are the Pauli matrices
operating on electron and nuclear spins. That is

\begin{equation}
X = \sigma_x, Y=\sigma_y, Z=\sigma_z.
\end{equation}

Operations which may be performed on any system are governed by the
Hamiltonian of the system. We now describe the effective spin
Hamiltonian for two adjacent qubits of the Kane quantum computer and
give a short physical motivation for each term which makes up the
overall Hamiltonian:

\begin{equation}
H = \sum_{i=1}^2 H_{B_i} + H_{A_i} + H_{J} + H_{ac_i}. 
	\label{eqn:sysHamiltonian}
\end{equation}
where the summation is over each donor atom in the system, $i$.

Under typical operating conditions, a constant magnetic field $B$ will
be applied to the entire system, perpendicular to the surface. This
contributes Zeeman energies to the Hamiltonian:

\begin{equation}
H_B = - g_n \mu_n B Z_n + \mu_B B Z_e.
\end{equation}
A typical value for the Kane quantum computer of $B = 2.0 \mathrm{T}$
gives Zeeman energy for the electrons of $\mu_B B \approx 0.116
\mathrm{meV}$, and for the nucleus $g_n \mu_n B \approx 7.1 \times
10^{-5} \mathrm{meV}$.

The hyperfine interaction couples between nuclear and electronic
spin. The contribution of the hyperfine interaction to the Hamiltonian
is

\begin{equation}
H_A = A \sigma_e \cdot \sigma_n,
\end{equation}
where strength, $A$, of the hyperfine interaction is proportional to
the value of the electron wave-function evaluated at the nucleus,

\begin{equation}
A = \frac{8\pi}{3} \mu_B g_n \mu_n |\psi(0)|^2.
\end{equation}

A typical strength for the hyperfine interaction is $A = 1.2 \times
10^{-4} \mathrm{meV}$. Charged A-Gates placed directly above each P
nucleus distort the shape of the electronic wavefunction thereby
reducing the strength of the hyperfine coupling. The nature of this
effect is under numerical investigation \cite{KG03}. For the
purposes of this paper we have assumed that it will be possible to
vary the hyperfine coupling by up to approximately $50\%$.

The exchange interaction couples adjacent electrons. Its contribution
to the Hamiltonian is:

\begin{equation}
H_J = J \sigma_{e_1} \cdot \sigma_{e_2},
\end{equation}
where $e_1$ and $e_2$ are two adjacent electrons. The magnitude, $J$,
of the exchange interaction depends on the overlap of adjacent
electronic wave-functions. J-Gates placed between nuclei distort both
electronic wavefunctions to increase or decrease the magnitude of this
interaction.  A typical value for the exchange energy is $4J = 0.124
\textrm{meV}$, and for the purposes of this paper we assume that it
will be possible to vary the magnitude of the exchange interaction
from $J=0$ to $J \approx 0.043 \mathrm{meV}$.

A rotating magnetic field, of strength $B_{ac}$ rotating at a
frequency of $\omega_{ac}$ can be applied, perpendicular to the
constant magnetic field, $B$.  The contribution of the rotating
magnetic field to the Hamiltonian is:

\begin{eqnarray}
H_{ac} &= - g_n \mu_n B_{ac} &\left[ X_n \cos(\omega_{ac}t) 
	+ Y_n \sin(\omega_{ac}t) \right] \nonumber\\
      &+ \mu_B B_{ac} & \left[X_e \cos(\omega_{ac}t) + Y_e
	\sin(\omega_{ac}t) \right], 
\end{eqnarray}
where the strength of the rotating magnetic field is envisioned to be $B_{ac}
\approx 0.0025 T$.

At an operating temperature of $T=100\mathrm{mK}$, the electrons are almost all
polarized by the magnetic field. That is:

\begin{equation}
\frac{n^{\uparrow}_e}{n^{\downarrow}_e} \approx 2.14 \times 10^{-12}.
\end{equation}

We assume that electrons are polarized in the $\ket{\downarrow}$
state, and use nuclear spin states as our computational basis.


\subsection{Z Rotations}

Single qubit rotations are required to implement the two qubit gates
described in this paper, as well as being essential for
universality. In fact, as we will see they contribute significantly to
the overall time and fidelity of each two qubit gate. It is therefore
important to consider the time required to implement $Z$, $X$ and $Y$
rotations.

In this subsection we describe how fast Z rotations may be performed
varying the voltage on the A-Gates only. A Z rotations is described by
the equation:

\begin{eqnarray}
R_z(\theta) = e^{i \frac{\theta}{2} Z}.
\end{eqnarray}
A Z gate (phase flip), may be implemented as a rotation. It is given
up to a global phase by:

\begin{equation}
Z = -i R_z\left(\pi\right).
\end{equation}

Under the influence of a constant magnetic field, $B$, to second order
in $A$ \cite{GM00}, each nuclei will undergo Larmor precession around
the Z axis, at frequency of
\begin{equation}
\hbar \omega_{l} = 2g_n \mu_n B+ 2A + \frac{2A^2}{\mu_B B + g_n\mu_nB}.
\label{eqn:Larmor}
\end{equation}
Z rotations may be performed by variation of the hyperfine interaction
from $A$ to $A_z$ giving a difference in rotation frequency of
\begin{eqnarray}
\hbar \omega_{z} = 2(A-A_z)  + \frac{2(A^2 - A_z^2)}{\mu_B B + g_n\mu_nB}.
\end{eqnarray}

Perturbing the hyperfine interaction for one of the atoms, and
allowing free evolution will rotate this atom with respect to the
rotation of the unperturbed atoms. The speed of single atom Z
rotations depends how much it is possible to vary the strength of the
hyperfine interaction, $A$. For numerical simulation we use the
typical values shown in Table \ref{tab:Zrot}.

\begin{table}
\begin{tabular}{|>{\centering}m{4cm}|>{\centering}m{1cm}|c|}
\hline \textbf{Description} & \textbf{Term} & \textbf{Value} \\
\hline
Unperturbed Hyperfine Interaction & $A$ & $0.1211 \times
10^{-3} \textrm{meV}$ \\ 
\hline 
Hyperfine Interaction During Z Rotation & $A_z$ & $0.0606 \times
10^{-3} \textrm{meV}$\\ 
\hline
\end{tabular}
\caption{Typical Parameters for a Z Rotation} \label{tab:Zrot}
\end{table}

Under these conditions a Z gate may be performed on a single nuclear
spin in approximately

\begin{equation}
t_Z \approx 0.021 \mathrm{\mu s}.
\end{equation}

These rotations occur in a rotating frame, that precesses around the
Z axis with a frequency equal to the Larmor frequency. We may have to
allow a small time of free evolution until nuclei that are not
affected by the Z rotation orientate themselves to their original
phase. The time required for this operation is less than

\begin{equation}
t_{F} \le 0.02 \mathrm{\mu s}
\end{equation}


\subsection{X and Y Rotations}

In this section we show how techniques, similar to those used in NMR
\cite{Bec00, Sli78, GM00}, may be used to implement X and Y
rotations. X and Y rotations are described by the equations:

\begin{eqnarray}
R_x(\theta) &=& e^{i \frac{\theta}{2} X},\\
R_y(\theta) &=& e^{i \frac{\theta}{2} Y}.
\end{eqnarray}

X and Y rotations are performed by application of a rotating magnetic
field, $B_{ac}$. The rotating magnetic field is resonant with the
Larmor precession frequency given in Eq.\ (\ref{eqn:Larmor}), that
is:

\begin{equation}
\omega_{ac} = \omega_{l}.
\end{equation}

In contrast to NMR, in the Kane proposal we have direct control over
the Larmor frequency of each individual $\mathrm{P}$ nucleus. By
reducing the hyperfine coupling for the atom we wish to target from
$A$ to $A_{x}$ we may apply an oscillating magnetic field that is only
resonant with the Larmor frequency of only one of the atoms. This
allows us to induce an X or Y rotation on an individual atom. To the
first order, the frequency of this rotation may be approximated by:

\begin{equation}
\hbar \omega_{x} = g_n \mu_n B_{ac} \left(1+\frac{A_x}{g_n\mu_nB}\right).
\end{equation}

The speed of an X rotation is directly proportional to the strength of
the rotating magnetic field, $B_{ac}$. As the strength of the rotating
magnetic field, $B_{ac}$ increases, the fidelity of the operation
decreases. The reason is that in frequency space the Full Width Half
Maximum (FWHM) of the transition excited by the rotating magnetic
field increases in proportional to $B_{ac}$. That is, as $B_{ac}$
increases we begin to excite non-resonant transitions. The larger
separation, in frequency space, between Larmor frequencies, the
smaller this systematic error. Since the Larmor precession frequency
depends on how much we are able to vary the hyperfine interaction,
$A$, it determines how strong we are able to make $B_{ac}$

%

For the purpose of simulation, the typical values shown in Table
\ref{tab:Xrot} for the unperturbed hyperfine interaction strength $A$,
the hyperfine interaction strength during the X rotation $A_x$,
applied magnetic field strength $B$, and rotating magnetic field
strength $B_{ac}$ were used.

\begin{table}
\begin{tabular}{|>{\centering}m{4cm}|>{\centering}m{1cm}|c|}
\hline \textbf{Description} & \textbf{Term} & \textbf{Value} \\
\hline 
Unperturbed Hyperfine Interaction & $A$ & $0.1211 \times
10^{-3} \textrm{meV}$ \\ 
\hline 
Hyperfine Interaction during X Rotation & $A_x$ & $0.0606 \times
10^{-3} \textrm{meV}$\\ 
\hline
Constant Magnetic Field Strength & $B$ & $2.000 \textrm{T}$ \\
\hline
Rotating Magnetic Field Strength & $B_{ac}$ & $0.0025 \textrm{T}$ \\ 
\hline
\end{tabular}
\caption{Typical Parameters for an X Rotation} \label{tab:Xrot}
\end{table}

Using these parameters this gives the overall time to perform an X
gate on a single qubit in approximately:

\begin{equation}
t_X \approx 6.4 \mu s.\\
\end{equation}

Any single qubit gate may be expressed as a product of X, Y and Z
rotations. Ideally, X and Y rotations should be minimized because Z
rotations may be performed much faster than X or Y rotations. For
example, a Hadamard gate may be expressed as a product of Z and X
rotations:

\begin{equation}
H = R_z \!\left(\frac{\pi}{2}\right) \  R_x\!\left(\frac{\pi}{2}\right) \
R_z\!\left(\frac{\pi}{2}\right).
\end{equation}
Thus, from the above discussion, Hadamard gate takes a time of approximately:

\begin{equation}
t_H \approx 3.2 \mu s.\\
\end{equation}


\subsection{Nuclear Spin Interaction}\label{sec:SecondOrder}

In this section we show the results of second order perturbation
theory to describe the interaction between two neighboring $P$
atoms. This interaction between nuclei is coupled by electron
interactions. We consider the case where the hyperfine couplings,
between each nucleus and its electron are equal, that is:
\begin{equation}
A = A_1 = A_2.
\end{equation}
We allow coupling between electrons, that is:
\begin{equation}
J > 0,
\end{equation}
but restrict ourselves to be far from an electronic energy level
crossing,
\begin{equation}
J \ll \frac{\mu_B B}{2}.
\end{equation}
Under these conditions electrons will remain in the polarized
$\ket{\downarrow \downarrow}$ ground state.

In this situation analysis has been performed using second order
perturbation theory \cite{GM00}. To second order in $A$, the energy
levels are:
\begin{eqnarray}
E_{\ket{11}} &=& -2\mu_BB +J + 2g_n\mu_nB + 2A,\\
E_{\ket{s_n}} &=& -2\mu_BB +J - \frac{2A^2}{\mu_BB+g_n\mu_nB},\\
E_{\ket{a_n}} &=& -2\mu_BB +J - \frac{2A^2}{\mu_BB+g_n\mu_nB - 2J},\\
E_{\ket{00}} &=& -2\mu_BB +J - 2g_n\mu_nB
 - 2A \nonumber \\ 
	& & \! - \frac{2A^2}{\mu_BB+g_n\mu_nB - 2J} -
\frac{2A^2}{\mu_BB+g_n\mu_nB}, \quad
\end{eqnarray}
where the symmetric $\ket{s_n}$ and anti-symmetric $\ket{a_n}$ energy
eigenstates are given by:
\begin{eqnarray}
\ket{s_n} &=& \frac{1}{\sqrt{2}} \left(\ket{10} + \ket{01} \right),\\
\ket{a_n} &=& \frac{1}{\sqrt{2}} \left(\ket{10} - \ket{01} \right).
\end{eqnarray}
Notice that the energies are symmetric around
\begin{eqnarray}
E_{0} 	&= -2\mu_BB & +J  \nonumber  - \frac{A^2}{\mu_BB+g_n\mu_nB -2J}\\
	 &  & - \frac{A^2}{\mu_BB+g_n\mu_nB}.
\end{eqnarray}
Since we are free to choose our zero point energy to be $E_0$ (or
equivalently ignore a global phase of a wavefunction, $\ket{\psi}$) we
may rewrite the second order approximation as:
\begin{eqnarray}
E_{\ket{\downarrow \downarrow}\ket{11}}  =& & \hbar \omega_B, \\
E_{\ket{\downarrow \downarrow}\ket{s_n}} =& & \hbar \omega_S, \\
E_{\ket{\downarrow \downarrow}\ket{a_n}} =& - & \hbar \omega_S, \\
E_{\ket{\downarrow \downarrow}\ket{00}}  =& - & \hbar \omega_B,
\end{eqnarray}
where $\omega_B$ and $\omega_S$ are given by:

\begin{eqnarray}
\hbar \omega_B &=& 2A + 2g_n \mu_n B + \nonumber \\ 
  & & \frac{A^2}{\mu_BB+g_n \mu_nB} + \frac{A^2}{\mu_BB+g_n \mu_nB -
  2J}, \label{eqn:omegaB}\\ 
\hbar \omega_S &=& \frac{A^2}{\mu_BB+g_n \mu_nB - 2J}
  - \frac{A^2}{\mu_BB+g_n \mu_nB}. \label{eqn:omegaS}
\end{eqnarray}
The reason for this representation of the energy will become clear in
the next section. Typical values were used during
numerical simulation of the interaction between nuclei are shown in
Table \ref{tab:Interaction}.

\begin{table}
\begin{tabular}{|>{\centering}m{3.5cm}|>{\centering}m{1cm}|c|}
\hline \textbf{Description} & \textbf{Term} & \textbf{Value}\\ 
\hline 
Hyperfine Interaction during Interaction  & $A_U$ &  $0.1197 \times
10^{-3} \  \textrm{meV}$ \\
\hline
Exchange Interaction during Interaction & $J_U$ & $0.0423 \  \textrm{meV}$\\
\hline
\end{tabular}
\caption{Typical Parameters during Interaction} \label{tab:Interaction}
\end{table}

\section{The Canonical Decomposition}\label{sec:Canonical}

In this section we describe the canonical decomposition, and describe
how this decomposition may be applied to the Kane quantum computer.


\subsection{Mathematical Description of Canonical Decomposition}

The canonical decomposition \cite{KC01, KBG00} decomposes any two
qubit unitary operator into a product of four single qubit unitaries
and one entangling unitary.

\begin{equation}
U = (V_1 \otimes V_2) \ U_{can} \ (W_1 \otimes W_2).
\end{equation}
where $V_1$, $V_2$, $W_1$ and $W_2$ are single qubit unitaries, and
$U_{can}$ is the two qubit interaction. The symbol $\otimes$
represents the tensor product of two matrices.

$U_{can}$ has a simple form involving only three parameters,
$\alpha_x$, $\alpha_y$ and $\alpha_z$:
\begin{equation}
U_{can} = e^{i \alpha_x X\otimes X} \ e^{i \alpha_y Y\otimes Y} \  e^{i
\alpha_z Z\otimes Z}.
\end{equation}
This purely non-local term is known as the \emph{interaction content}
of the gate. It is not difficult to show that each of the terms in the
interaction content, $e^{i \alpha_x X\otimes X}$, $e^{i \alpha_y
Y\otimes Y}$ and $e^{i \alpha_z Z\otimes Z}$, commute with each other.

\label{sec:controlRot}
Physically each of the terms $e^{i \alpha_x X\otimes X}$, $e^{i
\alpha_y Y\otimes Y}$, and $e^{i \alpha_z Z\otimes Z}$ correspond to
a type of controlled rotation. For example, following Ref.\ \cite{BDD+02}

\begin{eqnarray}
& & e^{i \alpha_z Z\otimes Z} \nonumber \\
&=& \cos{\alpha_z} I\otimes I + i \sin{\alpha_z} Z \otimes Z\nonumber \\ 
&=& \cos{\alpha_z} (\ket{0}\bra{0} + \ket{1}\bra{1}) \otimes I
\nonumber \\
& & + i \sin{\alpha_z} (\ket{0}\bra{0} - \ket{1}\bra{1}) \otimes Z \nonumber \\
&=& \ket{0}\bra{0} \otimes e^{  i \alpha_z Z} +
    \ket{1}\bra{1} \otimes e^{- i \alpha_z Z} \nonumber \\
&=& \left( I \otimes e^{i \alpha_z Z}\right) \  \left(\ket{0}\bra{0}
\otimes I \  + \ 
               		\ket{1}\bra{1} \otimes e^{- i 2\alpha_z Z}
\right). \label{eqn:CZ}
\end{eqnarray}

This shows that up to a single qubit rotation, $e^{i \alpha_z Z\otimes
Z}$ is equivalent to a controlled Z rotation. This holds true for the
other two terms. If we denote the eigenstates of X by

\begin{eqnarray}
X \ket{x_+} &=& +\ket{x_+},\\
X \ket{x_-} &=& -\ket{x_-},
\end{eqnarray}
then a similar analysis shows that

\begin{eqnarray}
& \left(I \otimes e^{-i\alpha_xX} \right) e^{i \alpha_x X\otimes X}\nonumber \\
&= \ket{x_+}\bra{x_+} \otimes I \  + \ 
    	 	\ket{x_-}\bra{x_-} \otimes e^{- i 2\alpha_x X},
\end{eqnarray}
and that

\begin{eqnarray}
&\left(I \otimes e^{-i\alpha_yY} \right) e^{i \alpha_y Y\otimes Y}
\nonumber \\
& = 	 	\ket{y_+}\bra{y_+} \otimes I \ + \ 
    	 	\ket{y_-}\bra{y_-} \otimes e^{- i 2\alpha_y Y}.
\end{eqnarray}
These operations are equivalent to controlled rotations in the $X$ and $Y$
directions respectively. For the first case, if the control qubit is
in the $\ket{x_-}$ state an X rotation is applied to the target qubit,
and not applied if the control qubit is in the $\ket{x_+}$
state. Similarly for Y.

Single qubit rotations, $V_1, V_2, W_1, W_2$ are possible on the Kane
quantum computing architecture, the remaining task is to specify the
pulse sequence for the purely entangling unitary
$U_{can}$. Fortunately this is always possible, as \emph{any}
interaction (with single qubit rotations) between the two nuclei is
sufficient \cite{DNB+01}. In fact, it is a relatively simple task to
use almost any interaction between qubits to generate any desired
operation.


\subsection{Calculation of the Interaction Content between Nuclei}

In this subsection we will see how it is possible to apply the
canonical decomposition to the Kane quantum computer. This is
important as this natural interaction of the system will be manipulated
by single qubit unitaries to find the pulse scheme of any two qubit
gate. The canonical decomposition provides a unique way of looking at
this interaction.

The interaction we will apply the canonical decomposition to is free
evolution of the configuration described in Sec.\
\ref{sec:SecondOrder}, using the results cited there from second order
perturbation theory. After a particular time of free evolution, our
system will have evolved according to unitary dynamics, which we may
decompose using the canonical decomposition:

\begin{equation}
U_{sys} = (V_1^{s} \otimes V_2^{s}) \ U_{can} ^{s} \ (W_1^{s} \otimes W_2^{s}),
\end{equation}
where the super-script `s' indicates a physical operation present in
our \emph{system}.

We wish to find the interaction content $U_{can}^{s}$ of this free
evolution. Systematic methods for doing this are given in \cite{KC01,
Has02, HVC02}. This is most easily done by noting any interaction
content, $U_{can}$ is diagonal in the so-called magic basis, otherwise
known as the Bell basis. This basis is:

\begin{eqnarray}
\ket{\Phi_1} &=& \frac{1}{\sqrt{2}}(\ket{00} + \ket{11}), \label{eqn:phi1}\\
\ket{\Phi_2} &=& \frac{-i}{\sqrt{2}}(\ket{00} - \ket{11}),\\
\ket{\Phi_3} &=& \frac{1}{\sqrt{2}}(\ket{01} - \ket{10}),\\
\ket{\Phi_4} &=& \frac{-i}{\sqrt{2}}(\ket{01} + \ket{10}). \label{eqn:phi4}
\end{eqnarray}

$\alpha_x$, $\alpha_y$ and $\alpha_z$ are related to the eigenvalues
$e^{i\lambda_1}$, $e^{i\lambda_2}$, $e^{i\lambda_3}$ and
$e^{i\lambda_4}$ of $U_{can}$. That is:

\begin{eqnarray}
\lambda_1 &=& +\alpha_x - \alpha_y + \alpha_z, \label{eqn:lambda1} \\
\lambda_2 &=& -\alpha_x + \alpha_y + \alpha_z,\\
\lambda_3 &=& -\alpha_x - \alpha_y - \alpha_z,\\
\lambda_4 &=& +\alpha_x + \alpha_y + \alpha_z. \label{eqn:lambda4}
\end{eqnarray}

It is possible to relate these eigenvalues to our system. After a
time $t$, each of the eigenstates of the system will have evolved
according the Schr\"odinger equation, which we may view as having
performed an operation $U_{sys}(t)$ on the system. As we showed in
Sec.\ \ref{sec:SecondOrder}:

\begin{eqnarray}
U_{sys}\ket{11} &=& e^{+i \theta_B} \ket{11}, \label{eqn:U11}\\ 
U_{sys}\ket{00} &=& e^{-i \theta_B} \ket{00},\\ 
U_{sys}\ket{s} &=& e^{+i \theta_S} \ket{s},\\
U_{sys}\ket{a} &=& e^{-i \theta_S} \ket{a}, \label{eqn:UA} 
\end{eqnarray}
where
\begin{eqnarray}
\theta_S &=& \omega_S t, \label{eqn:thetaS}\\
\theta_B &=& \omega_B t.
\end{eqnarray}
Applying Eqs.\ (\ref{eqn:U11}) -- (\ref{eqn:UA}) to
Eqs. (\ref{eqn:phi1}) -- (\ref{eqn:phi4}), we obtain:
\begin{eqnarray}
U_{sys}\ket{\Phi_1} &=& \cos(\theta_B)\ket{\Phi_1}-\sin(\theta_B)\ket{\Phi_2},\\
U_{sys}\ket{\Phi_2} &=& \cos(\theta_B)\ket{\Phi_2}+\sin(\theta_B)\ket{\Phi_1},\\
U_{sys}\ket{\Phi_3} &=& e^{-i \theta_S} \ket{\Phi_3},\\
U_{sys}\ket{\Phi_4} &=& e^{+i \theta_S} \ket{\Phi_4}.
\end{eqnarray}
This shows that in the magic basis, $U_{sys}$ is given by:
\begin{equation}
U_{sys} = \left[
	\begin{array}{cccc}	
	\cos(\theta_B) & \sin(\theta_B) & 0 & 0 \\
	- \sin(\theta_B) & \cos(\theta_B) & 0 & 0 \\
	0 & 0 & e^{-i\theta_S} & 0 \\
	0 & 0 & 0 & e^{i\theta_S} \\
	\end{array}
	\right].
\end{equation}

It is possible to find the eigenvalues $\lambda_1$, $\lambda_2$,
$\lambda_3$ and $\lambda_4$. We note that the eigenvalues of $U^TU$ in
the magic basis are given by

\begin{equation}
\lambda(U^TU) = \{e^{2i\lambda_1}, e^{2i\lambda_2}, e^{2i\lambda_3},
e^{2i\lambda_4} \}. \label{eqn:eigU}
\end{equation}
Calculation of the eigenvalues of $U_{sys}^TU_{sys}$ is easy since
$U_{sys}^TU_{sys}$ is already diagonal in this basis, with diagonal elements being $\{1, 1, e^{-2i \theta_S}, e^{2i \theta_S}\}$. Care must be
exercised at this point, because it is not clear which branch should
be used when taking the argument. In our case, as long as $0 \le
\theta_S \le \frac{\pi}{2}$ \cite{Has02} then

\begin{eqnarray}
\lambda_1 & = & 0, \\
\lambda_2 & = & 0, \\
\lambda_3 & = & -\theta_S, \\
\lambda_4 & = & +\theta_S.
\end{eqnarray}
Using Eqs. (\ref{eqn:lambda1}) -- (\ref{eqn:lambda4}) we may
solve for the coefficients $\alpha_x$, $\alpha_y$ and $\alpha_z$,
giving:
\begin{eqnarray}
\alpha_x^s &=& \frac{1}{2} \theta_S, \label{eqn:alphax} \\
\alpha_y^s &=& \frac{1}{2} \theta_S,\\
\alpha_z^s &=& 0  \label{eqn:alphaz}.
\end{eqnarray}

Single qubit rotations, $W_1^s, W_2^s, V_1^s, V_2^s$, induced are Z
rotations. Z rotations are fast, and may be canceled in comparatively
little time by single qubit Z rotations in the opposite direction:
\begin{equation}
(V_1^{s \dagger} \otimes V_2^{s \dagger}) \  U_{sys} \  (W_1^{s \dagger}
\otimes W_2^{s \dagger}) = U_{can}^{s}.
\end{equation}
For notational convenience we will now label the interaction content
of the system by an angle rather than by its time. The time for this
interaction may be calculated through Eqs. (\ref{eqn:alphax}) --
(\ref{eqn:alphaz}), (\ref{eqn:thetaS}) and
(\ref{eqn:omegaS}). Therefore, we write:
\begin{equation}
U_{can}^{s}(\phi) = e^{i \phi X \otimes X + i \phi Y \otimes
Y}, \label{eqn:ucan}
\end{equation}
where
\begin{equation}
\phi = \frac{1}{2} \theta_S.
\end{equation}

This analysis has been based on second order perturbation theory. As we
approach the electronic energy level crossing, this approach is no
longer valid. Close to this crossing numerical analysis shows the
eigenvalues are no longer symmetric which implies $\alpha_z^s$ becomes
non-zero. Unfortunately in this regime, we excite the system into
higher energy electronic configurations.

Given any two qubit gate, such as the CNOT gate, there are many
different possible choices of single qubit rotations and free
evolution that will implement a desired gate. Z rotations are faster
single qubit rotations than X and Y rotations, and therefore it is
desirable to minimize X and Y rotations in order to optimize the time
required, for any given two qubit gate.

\section{The CNOT and Controlled Z Gates}\label{sec:CNOT}

\subsection{Introduction}

The CNOT gate is a particularly often cited example of a two qubit
gate. CNOT and single qubit rotations are universal for quantum
computation \cite{DiV95}. Many implementations, including the
Kane proposal \cite{Kan98}, use this fact to demonstrate that they can,
in principle, perform any quantum algorithm. It is a member of the
so-called fault tolerant \cite{Sho96} set of gates, which are
universal for quantum computing, and are particularly important in
error correction. In this section we find a pulse scheme to implement
the CNOT gate on the Kane quantum computer.

Controlled Z rotations, sometimes known as controlled phase gates, are
some of the most important operations for implementing quantum
algorithms. In particular, one of the simplest ways to implement
quantum Fourier transformations (QFTs) uses multiple controlled Z
rotations (see for example \cite{NC01}). Single qubit rotations and
the controlled Z gate are, like the CNOT gate, universal for quantum
computation. Controlled Z rotations may be used in the construction of
controlled X and Y rotations. In this section we find a pulse scheme
to implement any controlled Z rotation on the Kane quantum computer.

Because these two gates have similar interaction contents we consider
them together. We will first show how to construct a controlled Z gate
of any angle, and use this gate directly to construct a CNOT gate.

A controlled Z rotation of angle $\theta$ is defined in the
computational basis by
\begin{equation}
U_{\Lambda Z}(\theta) = 
	\left[
	\begin{array}{cccc}
	1 & 0 & 0 & 0 \\
	0 & 1 & 0 & 0 \\
	0 & 0 & 1 & 0 \\
	0 & 0 & 0 & e^{i\theta}
	\end{array}
	\right].
\end{equation}
The canonical decomposition of the controlled Z rotation by an angle
$\theta$ has an interaction content consisting of:
\begin{eqnarray}
\alpha_x &=& 0, \label{eqn:CZx}\\
\alpha_y &=& 0,\\
\alpha_z &=& \frac{\theta}{2} \label{eqn:CZz}.
\end{eqnarray}
This interaction content may be found by using systematic methods
\cite{KC01, Has02, HVC02}. The controlled Z gate also requires a Z
rotation as described by Eq.\ (\ref{eqn:CZ}).

CNOT is defined in the computational basis by the matrix
\begin{equation}
U_{CNOT} = 
	\left[
	\begin{array}{cccc}
	1 & 0 & 0 & 0 \\
	0 & 1 & 0 & 0 \\
	0 & 0 & 0 & 1 \\
	0 & 0 & 1 & 0
	\end{array}
	\right].
\end{equation}
The canonical decomposition of CNOT has an interaction content with
angles of
\begin{eqnarray}
\alpha_x &=& 0, \\
\alpha_y &=& 0, \\
\alpha_z &=& \frac{\pi}{4}.
\end{eqnarray}

Since the CNOT and Controlled Z gates are both types of controlled
rotation similar to those described in Sec.\ \ref{sec:controlRot}, it
is not a surprise that they have a similar interaction content. In
fact, Control Z gates (that is, a controlled Z rotation by an angle of
$\pi$) and CNOT gates have an identical interaction content, and are
therefore equivalent up to single qubit rotations. A CNOT gate may be
constructed from a Control Z gate conjugated by $I \otimes H$.

\subsection{The Construction}

Our first task in finding a suitable pulse scheme for the controlled Z
rotation to find a pulse scheme which implements the interaction
content [Eqs. (\ref{eqn:CZx}) -- (\ref{eqn:CZz})] of the
controlled Z rotation. Techniques have direct analogues in NMR
\cite{Bec00, Sli78}.

The first technique \cite{BDD+02} is to conjugate by
$I\otimes X$, $I\otimes Y$, or $I\otimes Z$ to change the sign of two
of these parameters. For example:
\begin{eqnarray}
 (I \otimes Z) &e^{i \alpha_x X\otimes X + i \alpha_y Y\otimes Y + 
	       i \alpha_z Z\otimes Z} & (I \otimes Z) \nonumber\\
	=&  e^{-i \alpha_x X\otimes X - i \alpha_y
		Y\otimes Y + i \alpha_z Z\otimes Z}.
\end{eqnarray}
This can be useful, because it allows us to exactly cancel every
controlled rotation except one:
\begin{eqnarray}
(I \otimes Z) \ U_{can} \ (I \otimes Z)\ U_{can} = e^{i 2\alpha_z
Z\otimes Z}. \label{eqn:refocus}
\end{eqnarray}
In our case, however, it turns out that $\alpha_z^s = 0$. In order to
reorder the parameters a useful technique is to conjugate by
Hadamards \cite{BDD+02}. This is one of only several choices of single
qubit rotations which reorder the parameters. In this case the order
of the parameters is:
\begin{eqnarray}
& (H \otimes H) \  e^{i \alpha_x X\otimes X + i \alpha_y Y\otimes Y + 
	       i \alpha_z Z\otimes Z} \  (H \otimes H) \nonumber\\
	&=  e^{i \alpha_z X\otimes X + i \alpha_y
		Y\otimes Y + i \alpha_x Z\otimes Z}. \label{eqn:rearrange}
\end{eqnarray}
Combining these two techniques gives the following construction
\begin{eqnarray}
& & e^{i \theta Z \otimes Z} \nonumber \\ 
&=& (Z \otimes I) \ (H \otimes
H) \ U_{can}^s\left(\frac{\theta}{2}\right) \ (H \otimes H) \ (Z \otimes I)
\nonumber \\ 
& & \quad (H \otimes H) \ U_{can}^s\left(\frac{\theta}{2}\right) \ (H \otimes
H). \label{eqn:CZtheta}
\end{eqnarray}
To find the final construction, several one qubit optimizations were
made by combining adjacent single qubit rotations and using the
identities:
\begin{eqnarray}
HZH &=& X,\\ 
HH &=& I.
\end{eqnarray}

Operations may be performed in parallel. For example, performing
identical X or Y rotations on separate nuclei is a natural operation
of the system, because magnetic fields are applied
globally. Performing operations in parallel is faster, and also higher
fidelity than performing them one at a time.

The construction of the controlled Z rotation is shown in Fig.
\ref{fig:czDia}. In this circuit the single qubit rotations specified
in Eq.\ (\ref{eqn:CZ}) have been included. The period of interaction
between nuclei may be increased or decreased to produce controlled
rotations by any angle, $\theta$, as specified in Eqs.
(\ref{eqn:CZtheta}), (\ref{eqn:ucan}), (\ref{eqn:thetaS}) and
(\ref{eqn:omegaS}).

\begin{center}
\begin{figure}[ht!]
\scalebox{0.85}{


\unitlength 0.42mm

\begin{picture}(248, 86)

\linethickness{0.1mm}
\put(0, 24) {\line(1,0) {10}}
\put(0, 62) {\line(1,0) {10}}
\linethickness{0.4mm}
\put(10, 10){\framebox(28, 28){$H$}}
\linethickness{0.4mm}
\put(10, 48){\framebox(28, 28){$H$}}
\linethickness{0.1mm}
\linethickness{0.1mm}
\put(38, 24) {\line(1,0) {10}}
\put(38, 62) {\line(1,0) {10}}
\linethickness{0.4mm}
\put(48, 10){\framebox(28, 66){$U\left(\frac{\pi}{8}\right)$}}
\linethickness{0.1mm}
\linethickness{0.1mm}
\put(76, 24) {\line(1,0) {10}}
\put(76, 62) {\line(1,0) {10}}
\linethickness{0.4mm}
\put(86, 48){\framebox(28, 28){$X$}}
\linethickness{0.1mm}
\put(86, 24) {\line(1,0) {28}}
\linethickness{0.1mm}
\put(114, 24) {\line(1,0) {10}}
\put(114, 62) {\line(1,0) {10}}
\linethickness{0.4mm}
\put(124, 10){\framebox(28, 66){$U\left(\frac{\pi}{8}\right)$}}
\linethickness{0.1mm}
\linethickness{0.1mm}
\put(152, 24) {\line(1,0) {10}}
\put(152, 62) {\line(1,0) {10}}
\linethickness{0.4mm}
\put(162, 10){\framebox(28, 28){$H$}}
\linethickness{0.4mm}
\put(162, 48){\framebox(28, 28){$H$}}
\linethickness{0.1mm}
\linethickness{0.1mm}
\put(190, 24) {\line(1,0) {10}}
\put(190, 62) {\line(1,0) {10}}
\linethickness{0.4mm}
\put(200, 10){\framebox(28, 28){$R_{z} \left(\frac{\pi}{2}\right)$}}
\linethickness{0.4mm}
\put(200, 48){\framebox(28, 28){$R_{z} \left(\frac{-\pi}{2}\right)$}}
\linethickness{0.1mm}
\linethickness{0.1mm}
\put(228, 24) {\line(1,0) {10}}
\put(228, 62) {\line(1,0) {10}}

\end{picture}

}
\caption{Circuit Diagram for Controlled Z Pulse Sequence} \label{fig:czDia}
\end{figure}
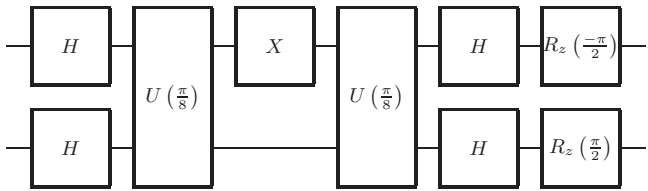
\end{center}
Our task of constructing a CNOT is now comparatively simple. We note
that a CNOT gate has the same interaction term as the controlled Z
(controlled phase) operation. These gates are therefore equivalent up
to local operations.

\begin{figure}[ht!]
\begin{center}
\scalebox{0.85}{


\unitlength 0.42mm

\begin{picture}(248, 86)

\linethickness{0.1mm}
\put(0, 24) {\line(1,0) {10}}
\put(0, 62) {\line(1,0) {10}}
\linethickness{0.4mm}
\put(10, 48){\framebox(28, 28){$H$}}
\linethickness{0.4mm}
\put(10, 10){\framebox(28, 28){$R_{x} \left(\frac{\pi}{2}\right)$}}
\linethickness{0.1mm}
\linethickness{0.1mm}
\put(38, 24) {\line(1,0) {10}}
\put(38, 62) {\line(1,0) {10}}
\linethickness{0.4mm}
\put(48, 10){\framebox(28, 66){$U\left(\frac{\pi}{8}\right)$}}
\linethickness{0.1mm}
\linethickness{0.1mm}
\put(76, 24) {\line(1,0) {10}}
\put(76, 62) {\line(1,0) {10}}
\linethickness{0.4mm}
\put(86, 48){\framebox(28, 28){$X$}}
\linethickness{0.1mm}
\put(86, 24) {\line(1,0) {28}}
\linethickness{0.1mm}
\put(114, 24) {\line(1,0) {10}}
\put(114, 62) {\line(1,0) {10}}
\linethickness{0.4mm}
\put(124, 10){\framebox(28, 66){$U\left(\frac{\pi}{8}\right)$}}
\linethickness{0.1mm}
\linethickness{0.1mm}
\put(152, 24) {\line(1,0) {10}}
\put(152, 62) {\line(1,0) {10}}
\linethickness{0.4mm}
\put(162, 48){\framebox(28, 28){$H$}}
\linethickness{0.1mm}
\put(162, 24) {\line(1,0) {28}}
\linethickness{0.1mm}
\put(190, 24) {\line(1,0) {10}}
\put(190, 62) {\line(1,0) {10}}
\linethickness{0.4mm}
\put(200, 48){\framebox(28, 28){$R_{z} \left(\frac{-\pi}{2}\right)$}}
\linethickness{0.1mm}
\put(200, 24) {\line(1,0) {28}}
\linethickness{0.1mm}
\put(228, 24) {\line(1,0) {10}}
\put(228, 62) {\line(1,0) {10}}

\end{picture}

}
\caption{Circuit diagram for CNOT pulse sequence} \label{fig:CNOTDia}
\end{center}
\end{figure}
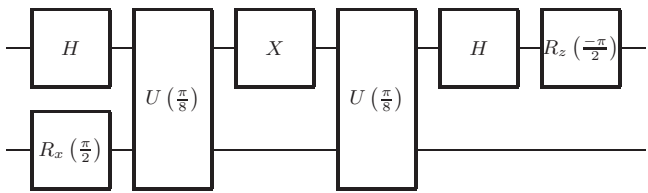

Conjugation by $I\otimes H$ will turn a controlled Z operation into a
CNOT gate. Using some simple one qubit identities to simplify the
rotations at the beginning and end of the pulse sequences we arrive at
the decomposition illustrated in the circuit diagram shown in Fig.
\ref{fig:CNOTDia}.

\subsection{Time and Fidelity}

Throughout the paper we define fidelity as

\begin{equation}
F(\ket{\psi}, \ket{\psi_{0}}) = | \braket{\psi}{\psi_0} | ^2,
\end{equation}
with $\ket{\psi}$ being the actual state obtained from evolution, and
$\ket{\psi_0}$ being the state which is desired. We define the error
in terms of the fidelity as

\begin{equation}
E = \max_{\ket{\psi}} \left[ 1-F(\ket{\psi},
\ket{\psi_{0}}) \right]. \label{eqn:Error}
\end{equation}
where the maximumization is performed over the output of all the
computational basis states, $\ket{\psi}$.

\begin{figure*}
\subfigure[$\ket{00}$ initial state] { \includegraphics[width=8cm]{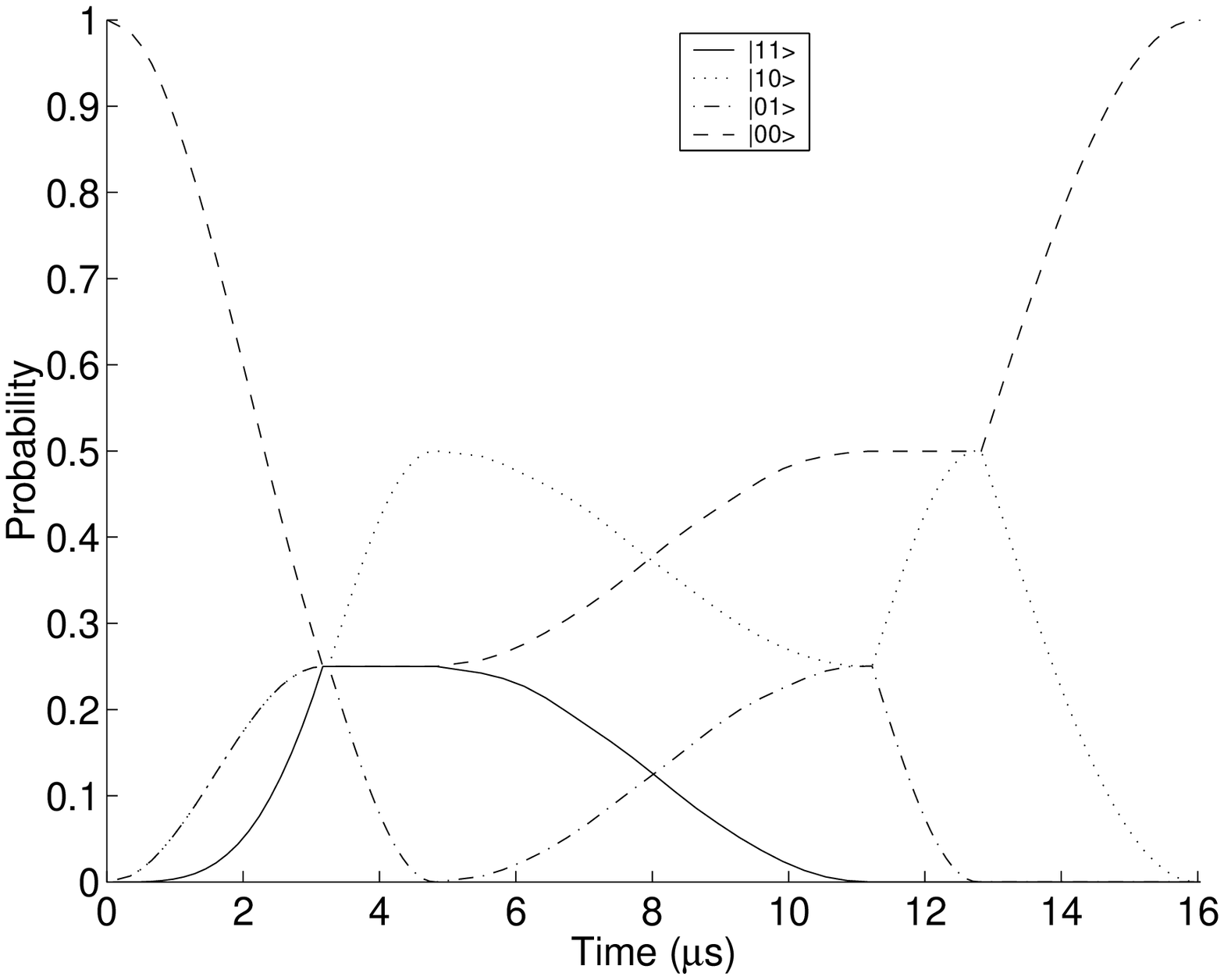} }
\subfigure[$\ket{01}$ initial state] { \includegraphics[width=8cm]{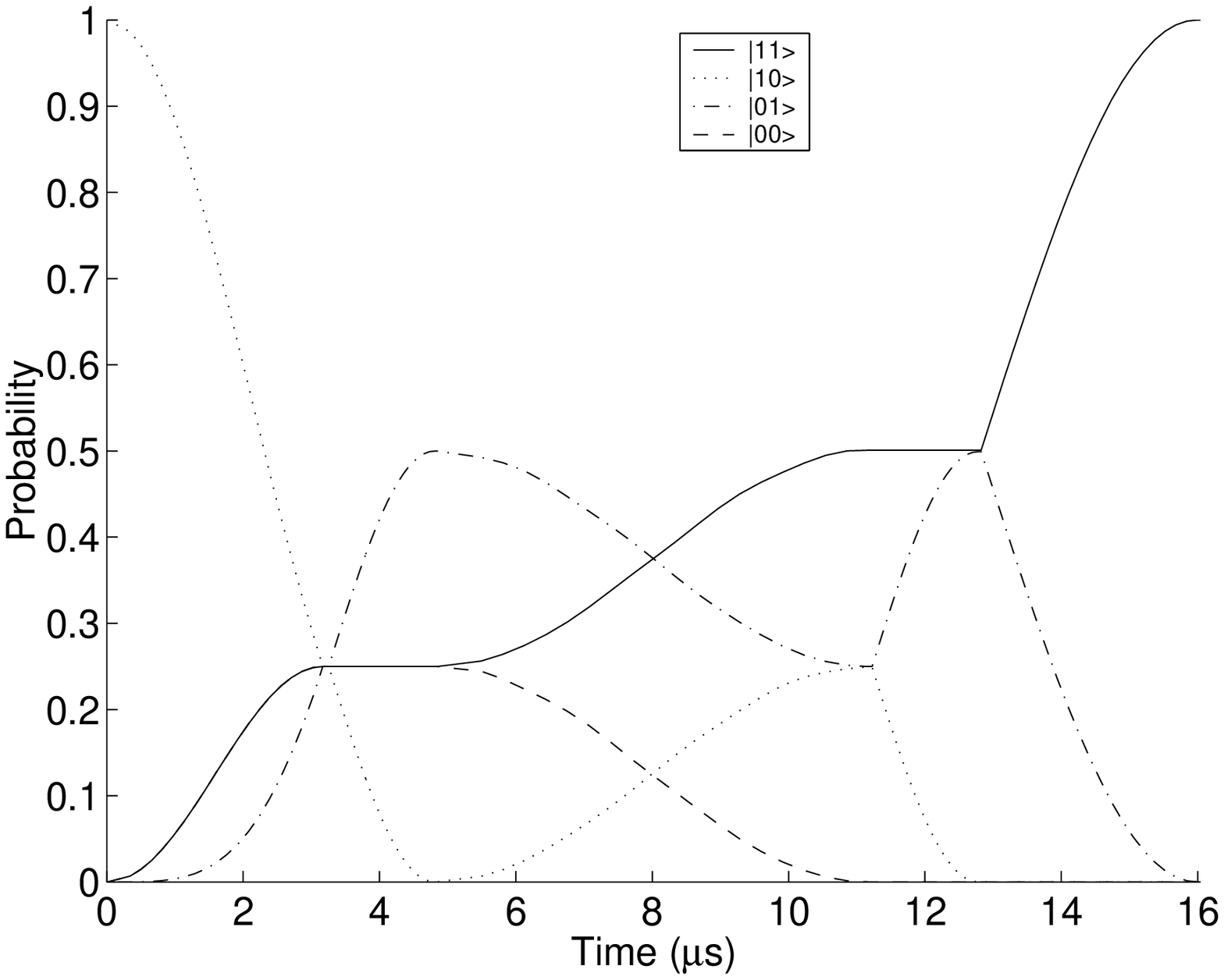} }
\subfigure[$\ket{10}$ initial state] { \includegraphics[width=8cm]{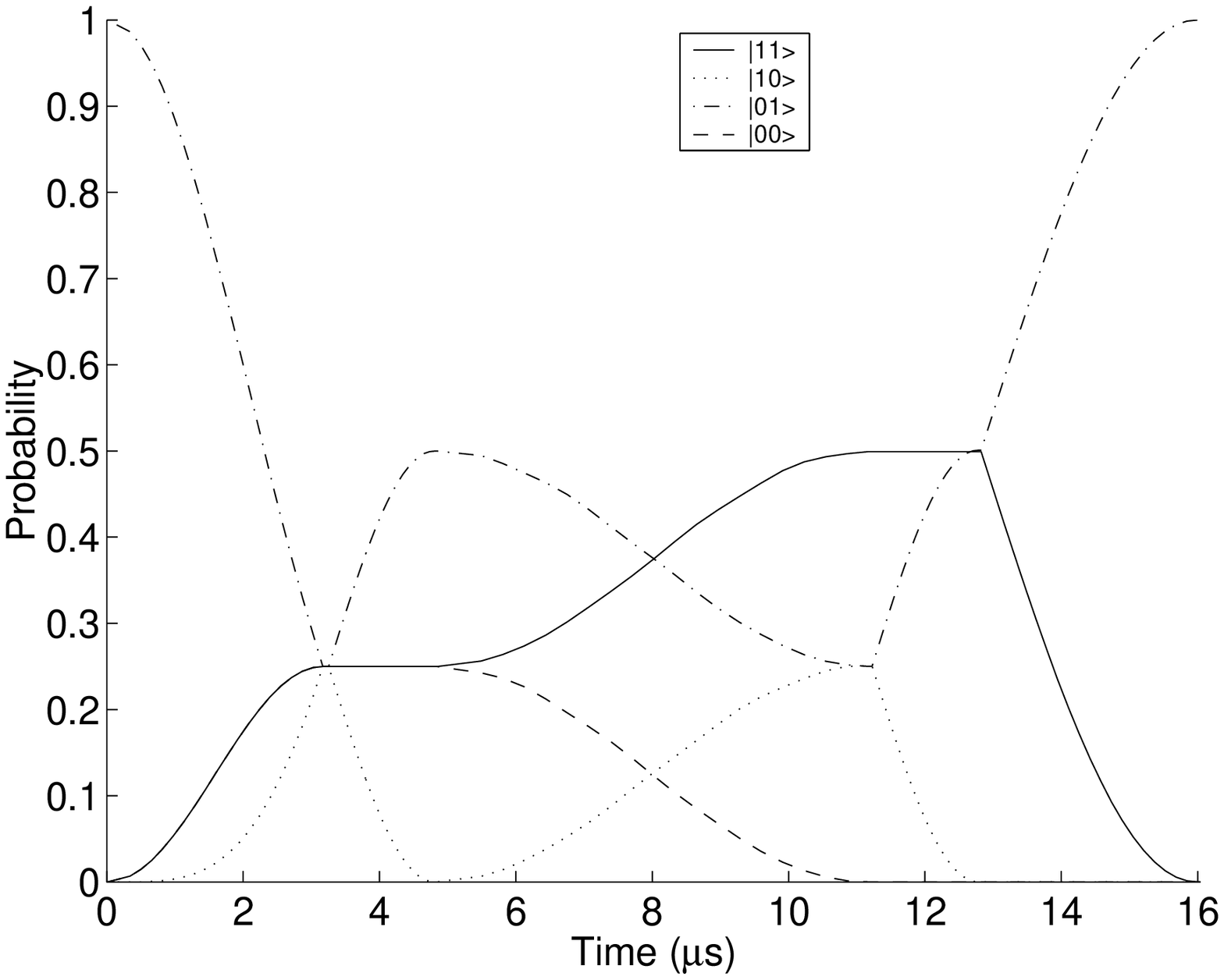} }
\subfigure[$\ket{11}$ initial state] { \includegraphics[width=8cm]{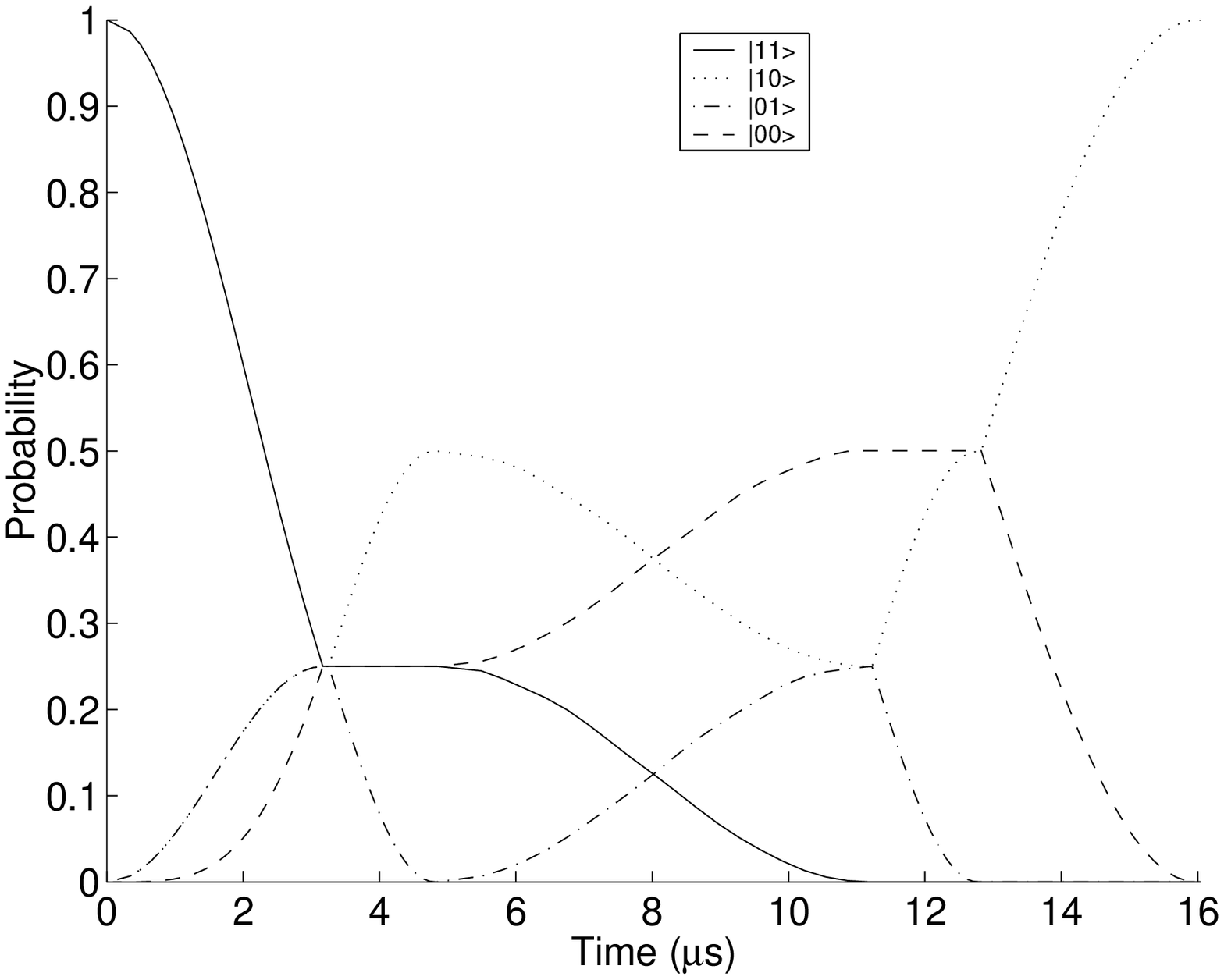} }
\caption{Numerical simulation of the CNOT gate showing different
initial conditions} \label{fig:cnotTrans}
\end{figure*}

Numerical simulations were carried out by numerically integrating
Schrodinger's equation for the Hamiltonian of the system, Eq.
(\ref{eqn:sysHamiltonian}). The results of this numerical simulation
for the pulse sequence of the CNOT gate are shown in
Fig. \ref{fig:cnotTrans}. These graphs show each of the states and the
transitions which are made. In these figures it is possible to see the
evolution of each of the four computational basis states. The control
qubit is the second qubit and the target qubit is the first qubit.

According to the numerical results, a full Controlled Z gate takes a
total time of $16.1\mu s$ and has an error of approximately $4 \times
10^{-5}$. Similarly we find the CNOT gate takes a total time
of $16.0 \mathrm{\mu s}$. The time required for this gate can be
grouped as shown in Table \ref{tab:breakdown}.

\begin{table}
\begin{tabular}{|c|r|}
\hline
\textbf{Description} & \textbf{Time}\\
\hline
X rotations &       	12.6 $\mathrm{\mu s}$ \\
Z rotations &       	0.2  $\mathrm{\mu s}$ \\
2 qubit interaction & 	3.2  $\mathrm{\mu s}$ \\
\hline
\textbf{Total} & 	16.0 $\mathrm{\mu s}$\\
\hline
\end{tabular}
\caption{Time for CNOT Gate} \label{tab:breakdown}
\end{table}

$X$ and $Y$ rotations make up the majority of the time taken to
implement the controlled Z and CNOT gates. In the CNOT gate, only $3.2
\mathrm{\mu s}$ is spent implementing the entangling part of the gate,
whereas $12.6 \mathrm{\mu s}$ is required to implement the $X$ and $Y$
rotations.

We can see via simulation that the systematic error in the CNOT gate
is approximately $4 \times 10^{-5}$. Some of this error will be due to
errors during simulation, and breakdown of the second order
approximation. A large part of the error, particularly if the
hyperfine interaction may not be varied very much, is due to X
rotations where unintended non-resonant transitions are excited along
with the intended rotation.

\section{The Swap and Square Root of Swap Gates}\label{sec:Swap}

\subsection{Introduction}

\begin{figure*}
\begin{center}
\scalebox{0.9}{


\unitlength 0.42mm

\begin{picture}(265, 80)

\linethickness{0.1mm}
\put(0, 23) {\line(1,0) {10}}
\put(0, 58) {\line(1,0) {10}}
\linethickness{0.4mm}
\put(10, 10){\framebox(25, 25){$H$}}
\linethickness{0.4mm}
\put(10, 45){\framebox(25, 25){$H$}}
\linethickness{0.1mm}
\linethickness{0.1mm}
\put(35, 23) {\line(1,0) {10}}
\put(35, 58) {\line(1,0) {10}}
\linethickness{0.4mm}
\put(45, 10){\framebox(25, 60){$U\left(\frac{\pi}{8}\right)$}}
\linethickness{0.1mm}
\linethickness{0.1mm}
\put(70, 23) {\line(1,0) {10}}
\put(70, 58) {\line(1,0) {10}}
\linethickness{0.4mm}
\put(80, 45){\framebox(25, 25){$X$}}
\linethickness{0.1mm}
\put(80, 23) {\line(1,0) {25}}
\linethickness{0.1mm}
\put(105, 23) {\line(1,0) {10}}
\put(105, 58) {\line(1,0) {10}}
\linethickness{0.4mm}
\put(115, 10){\framebox(25, 60){$U\left(\frac{\pi}{8}\right)$}}
\linethickness{0.1mm}
\linethickness{0.1mm}
\put(140, 23) {\line(1,0) {10}}
\put(140, 58) {\line(1,0) {10}}
\linethickness{0.4mm}
\put(150, 10){\framebox(25, 25){$H$}}
\linethickness{0.4mm}
\put(150, 45){\framebox(25, 25){$H$}}
\linethickness{0.1mm}
\linethickness{0.1mm}
\put(175, 23) {\line(1,0) {10}}
\put(175, 58) {\line(1,0) {10}}
\linethickness{0.4mm}
\put(185, 45){\framebox(25, 25){$Z$}}
\linethickness{0.1mm}
\put(185, 23) {\line(1,0) {25}}
\linethickness{0.1mm}
\put(210, 23) {\line(1,0) {10}}
\put(210, 58) {\line(1,0) {10}}
\linethickness{0.4mm}
\put(220, 10){\framebox(25, 60){$U\left(\frac{\pi}{4}\right)$}}
\linethickness{0.1mm}
\linethickness{0.1mm}
\put(245, 23) {\line(1,0) {10}}
\put(245, 58) {\line(1,0) {10}}

\end{picture}

}
\caption{Circuit Diagram for the Swap gate pulse sequence}
\label{fig:swapCircuitCCT}
\end{center}
\end{figure*}
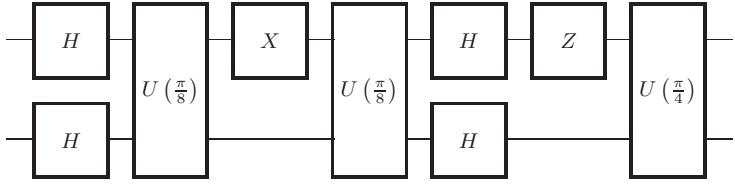

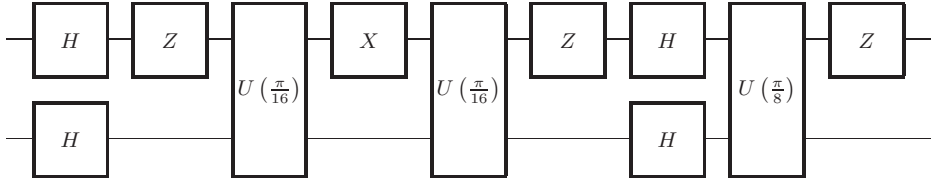
\begin{figure*}
\begin{center}
\scalebox{0.9}{


\unitlength 0.42mm

\begin{picture}(335, 80)

\linethickness{0.1mm}
\put(0, 23) {\line(1,0) {10}}
\put(0, 58) {\line(1,0) {10}}
\linethickness{0.4mm}
\put(10, 10){\framebox(25, 25){$H$}}
\linethickness{0.4mm}
\put(10, 45){\framebox(25, 25){$H$}}
\linethickness{0.1mm}
\linethickness{0.1mm}
\put(35, 23) {\line(1,0) {10}}
\put(35, 58) {\line(1,0) {10}}
\linethickness{0.4mm}
\put(45, 45){\framebox(25, 25){$Z$}}
\linethickness{0.1mm}
\put(45, 23) {\line(1,0) {25}}
\linethickness{0.1mm}
\put(70, 23) {\line(1,0) {10}}
\put(70, 58) {\line(1,0) {10}}
\linethickness{0.4mm}
\put(80, 10){\framebox(25, 60){$U\left(\frac{\pi}{16}\right)$}}
\linethickness{0.1mm}
\linethickness{0.1mm}
\put(105, 23) {\line(1,0) {10}}
\put(105, 58) {\line(1,0) {10}}
\linethickness{0.4mm}
\put(115, 45){\framebox(25, 25){$X$}}
\linethickness{0.1mm}
\put(115, 23) {\line(1,0) {25}}
\linethickness{0.1mm}
\put(140, 23) {\line(1,0) {10}}
\put(140, 58) {\line(1,0) {10}}
\linethickness{0.4mm}
\put(150, 10){\framebox(25, 60){$U\left(\frac{\pi}{16}\right)$}}
\linethickness{0.1mm}
\linethickness{0.1mm}
\put(175, 23) {\line(1,0) {10}}
\put(175, 58) {\line(1,0) {10}}
\linethickness{0.4mm}
\put(185, 45){\framebox(25, 25){$Z$}}
\linethickness{0.1mm}
\put(185, 23) {\line(1,0) {25}}
\linethickness{0.1mm}
\put(210, 23) {\line(1,0) {10}}
\put(210, 58) {\line(1,0) {10}}
\linethickness{0.4mm}
\put(220, 45){\framebox(25, 25){$H$}}
\linethickness{0.4mm}
\put(220, 10){\framebox(25, 25){$H$}}
\linethickness{0.1mm}
\linethickness{0.1mm}
\put(245, 23) {\line(1,0) {10}}
\put(245, 58) {\line(1,0) {10}}
\linethickness{0.4mm}
\put(255, 10){\framebox(25, 60){$U\left(\frac{\pi}{8}\right)$}}
\linethickness{0.1mm}
\linethickness{0.1mm}
\put(280, 23) {\line(1,0) {10}}
\put(280, 58) {\line(1,0) {10}}
\linethickness{0.4mm}
\put(290, 45){\framebox(25, 25){$Z$}}
\linethickness{0.1mm}
\put(290, 23) {\line(1,0) {25}}
\linethickness{0.1mm}
\put(315, 23) {\line(1,0) {10}}
\put(315, 58) {\line(1,0) {10}}

\end{picture}

}
\caption{The Circuit Diagram for the Square Root of Swap Pulse
Sequence} \label{fig:sqrtSwapCCT}
\end{center}
\end{figure*}

One of the most important gates for the Kane quantum computer is
envisioned to be the swap gate. This is because, in the Kane proposal,
only nearest neighbor interactions are allowed.  This gate swaps the
quantum state of two qubits. By using the swap gate it is possible to
swap qubits until they are nearest neighbors, interact them, and then
swap them back again. Having an efficient method to interact qubits
which are not adjacent to each other is therefore important, and the
swap gate, with its high level of information transfer, is one
possible method of achieving this.

The square root of swap gate has been suggested for the quantum dot
spin based quantum computer architecture \cite{LD98}, where it is a
particularly natural operation. In our system it is not such a natural
operation, but that does not mean that we cannot construct it. Like
the CNOT gate, the square root of swap (together with single qubit
rotations) is universal for quantum computation. In this section we
find a pulse sequence to implement both the swap and the square root
of swap gates on the Kane quantum computer architecture.

The swap gate is defined in the computational basis by:

\begin{equation}
U_{Swap} = 
	\left[
	\begin{array}{cccc}
	1 & 0 & 0 & 0 \\
	0 & 0 & 1 & 0 \\
	0 & 1 & 0 & 0 \\
	0 & 0 & 0 & 1
	\end{array}
	\right]
\end{equation}

The canonical decomposition of the swap gate has an interaction content
with angles of:
\begin{eqnarray}
\alpha_x &=& \frac{\pi}{4}\\
\alpha_y &=& \frac{\pi}{4}\\
\alpha_z &=& \frac{\pi}{4}
\end{eqnarray}

The square root of swap gate is defined in the computational basis by:
\begin{equation}
U_{SS} = 
	\left[
	\begin{array}{cccc}
	1 & 0 & 0 & 0 \\
	0 & \frac{1}{2}(1+i) & \frac{1}{2}(1-i)  & 0 \\
	0 & \frac{1}{2}(1-i) & \frac{1}{2}(1+i) & 0 \\
	0 & 0 & 0 & 1
	\end{array}
	\right]
\end{equation}

The canonical decomposition of the square root of swap gate has an
interaction term consisting of:
\begin{eqnarray}
\alpha_x &=& -\frac{\pi}{8},\\
\alpha_y &=& -\frac{\pi}{8},\\
\alpha_z &=& -\frac{\pi}{8}.
\end{eqnarray}

Since the square root of swap and swap gates have essentially the same
interaction content, their constructions are very similar, and are
therefore considered together here.

\subsection{The Construction}

The easiest way to construct a swap gate is simply to use free
evolution to obtain the angles $\alpha_x$ and $\alpha_y$ which is
natural for our system. The only remaining term is the $\alpha_z$
term, which for our system will naturally be 0. We may obtain this
term by applying a pulse sequence similar to the Controlled Z rotation
as described in Sec.\ \ref{sec:CNOT}. The resulting construction swap
gate is shown in the diagram in Fig. \ref{fig:swapCircuitCCT}.

The interaction content of the square root of swap gate is exactly
half that of the swap gate, and it is negative. We use exactly the
same technique used to obtain the swap gate, only allowing the nuclei
to interact for exactly half the time. To make the terms negative we
conjugate by $Z \otimes I$. The construction of the square root of
swap gate obtained using this method is shown in
Fig. \ref{fig:sqrtSwapCCT}.

\subsection{Speed and Fidelity}

\begin{figure*}[ht!]
\subfigure[$\ket{00}$ initial state] { \includegraphics[width=8cm]{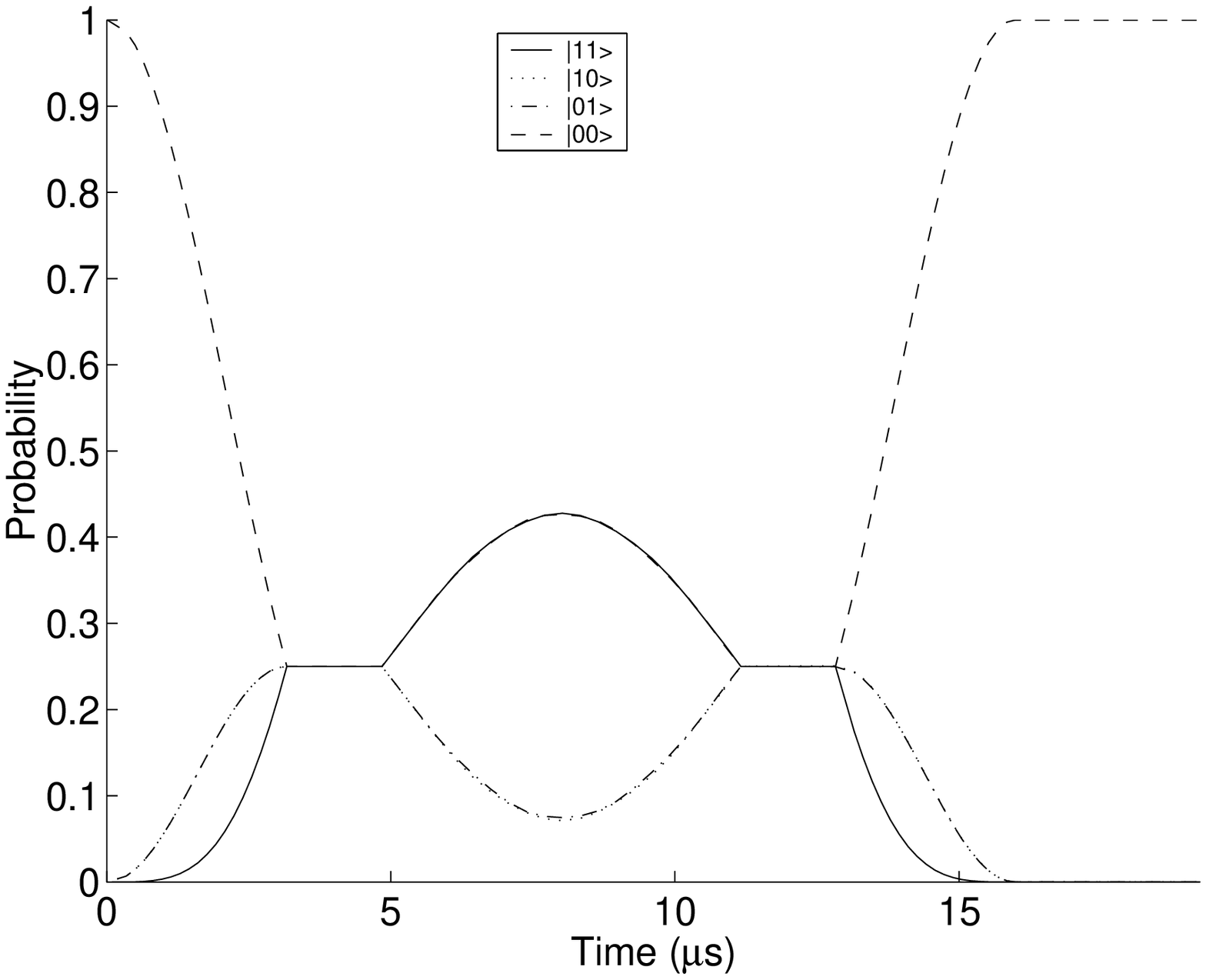} }
\subfigure[$\ket{01}$ initial state] { \includegraphics[width=8cm]{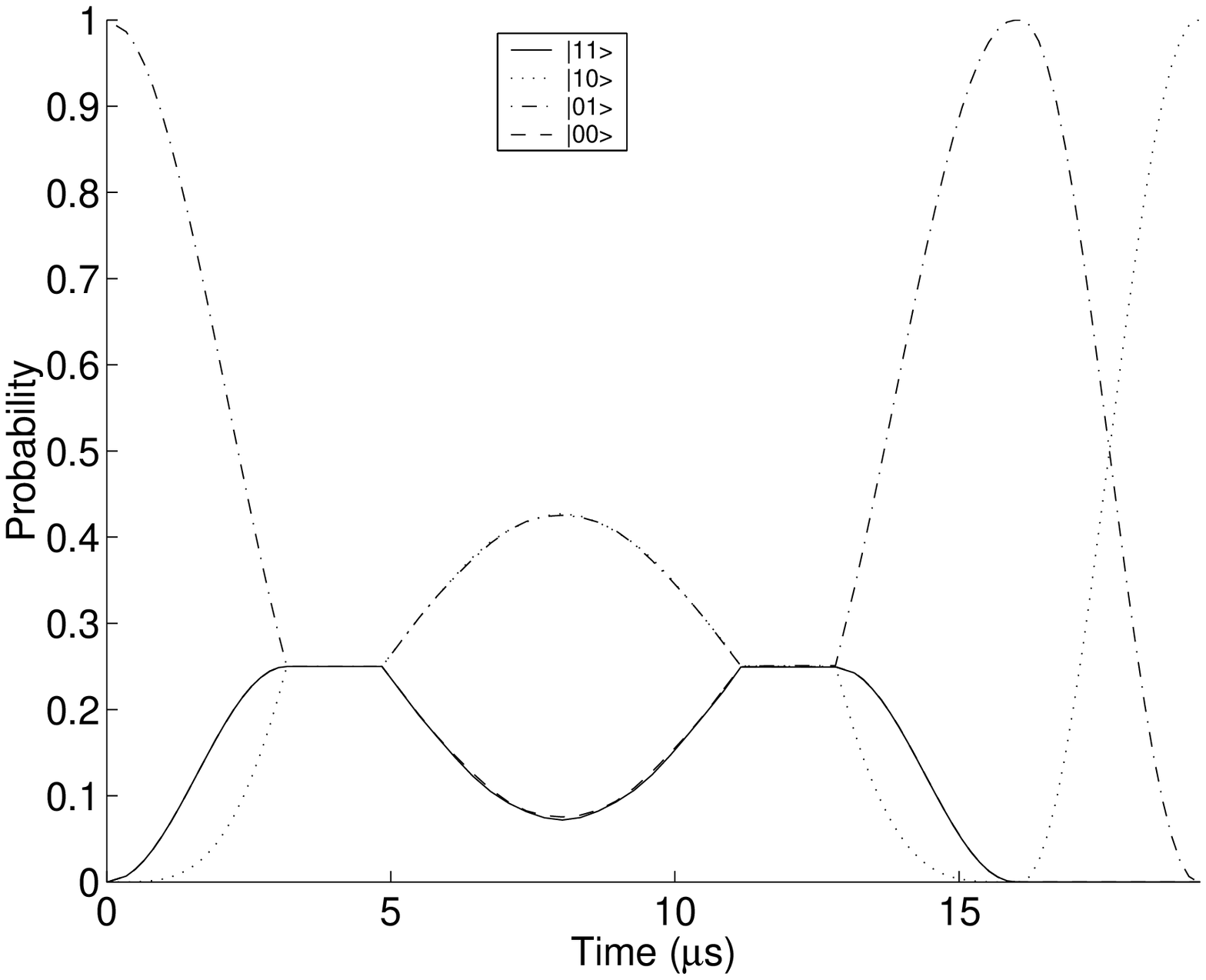} }
\subfigure[$\ket{10}$ initial state] { \includegraphics[width=8cm]{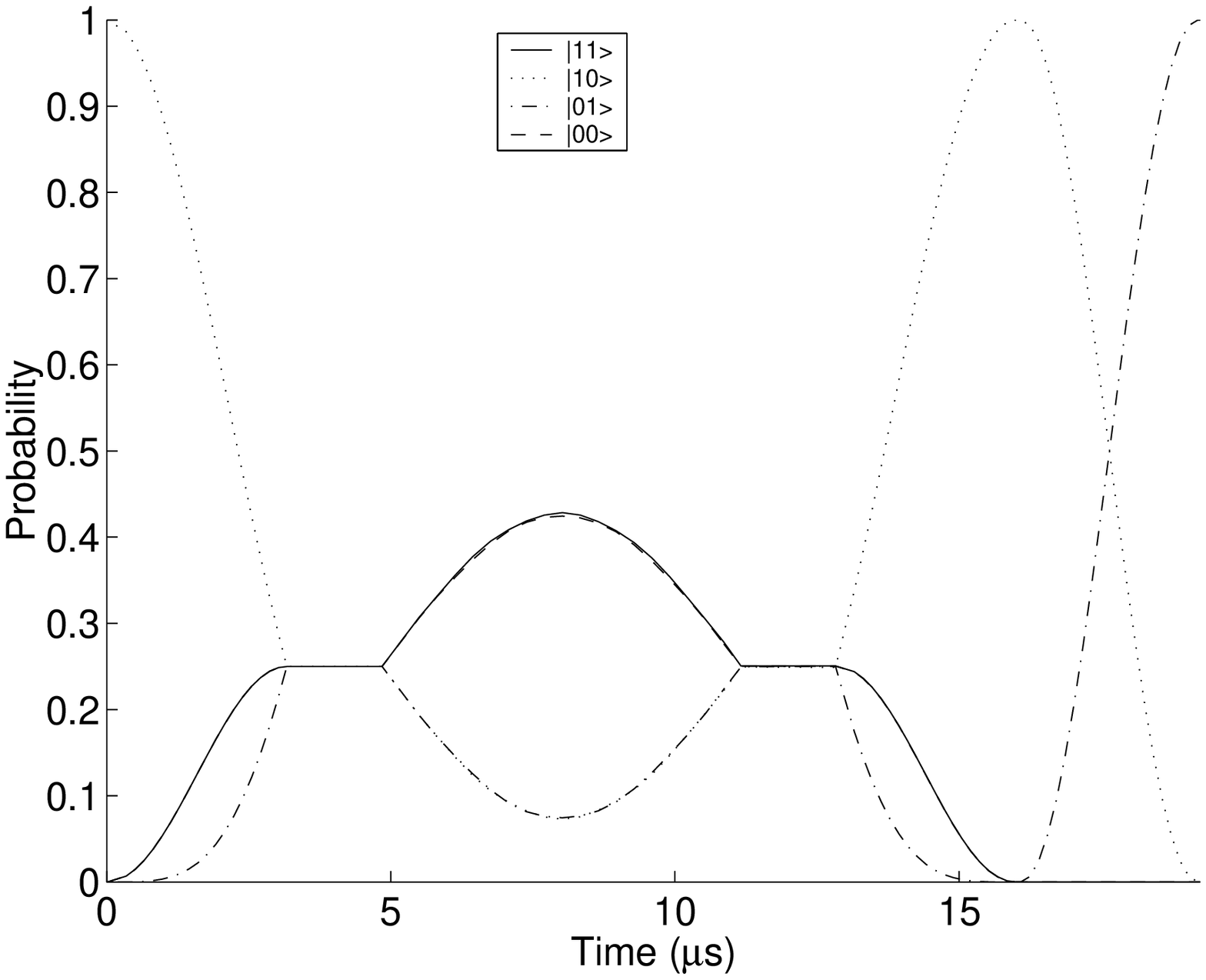} }
\subfigure[$\ket{11}$ initial state] { \includegraphics[width=8cm]{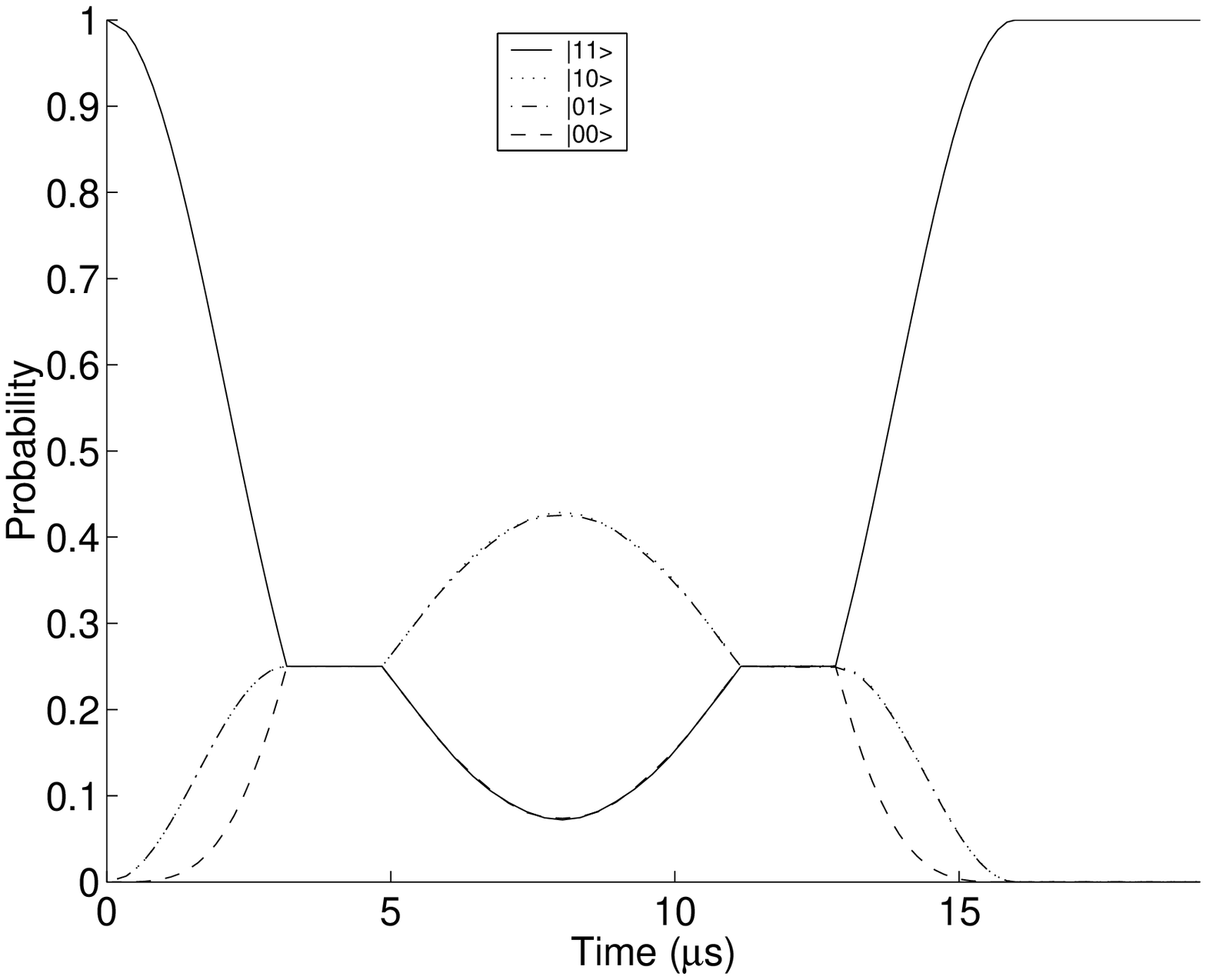} }
\caption{Numerical Simulation of the Swap Gate} \label{fig:swap}
\end{figure*}

The swap and square root of swap gates were simulated numerically. The
resulting transitions for the swap gate are shown in Fig.
\ref{fig:swap}. Similar results were obtained for the square root of
swap gate, not shown here.

The swap gate takes a total time of $19.2 \mu s$, and has a fidelity
of approximately $7 \times 10^{-5}$. The majority of time in this gate
is taken by X and Y rotations, which are also the major source of
error.

This is substantially faster than an existing suggestion for the swap
gate \cite{Wel01} of $192 \mu s$. It is also faster than using
three adiabatic CNOT gates, which would take approximately $78 \mu s$.

According to numerical simulation the square root of swap gate takes
$16.8 \mu s$ and has an error of approximately $5 \times
10^{-5}$. This is the first explicit proposal for the Kane quantum
computer for the square root of swap gate.

The square root of swap gate has been suggested in the context of
quantum computation for quantum dots \cite{LD98}. It is universal for
quantum computation and therefore can be used to construct a CNOT
gate. Unfortunately in this case, a CNOT constructed from the square
root of swap gate presented here would take approximately $40 \mu s$
which is much longer than the pulse sequence presented in this paper
for the CNOT gate.\\

\section{Conclusion}\label{sec:Conclusion}

We have shown how the canonical decomposition may be applied to the
Kane quantum computer. We found the canonical decomposition of a
natural operation of the computer, that is, free evolution with
hyperfine interactions equal and the exchange interaction non-zero. We
then used this interaction to form two qubit gates which may be applied
to the Kane quantum computer. These gates and their times and
fidelities are shown in Table \ref{tab:gateTimes}.

\begin{table}
\begin{tabular}{|c|r|r|}
\hline
\textbf{Gate} & \textbf{Time} & \textbf{Error} \\
\hline
CNOT 		& $16.0 \mu s$ 	& $4 \times 10^{-5}$\\ 
Swap 		& $19.2 \mu s$ 	& $7 \times 10^{-5}$\\
Square Root of Swap & $16.2\mu s$ & $5 \times 10^{-5}$\\
Controlled Z 	& $16.1 \mu s$ 	& $4 \times 10^{-5}$\\
\hline
\end{tabular}
\caption{Gate Times and Fidelities} \label{tab:gateTimes}
\end{table}

The majority of the time required to implement each of these two qubit
gates is used to implement single qubit rotations. Were we able to
perform these rotations faster and more accurately then the gates
presented here would also benefit. Another possible avenue of research
is to investigate the effect of decoherence on the system.

To our knowledge this is the fastest proposal for swap, square root of
swap, CNOT and controlled Z operations on the Kane quantum computer
architecture. We have shown how a representative set of two qubit
gates may be implemented on the Kane quantum computer. These methods
may prove particularly powerful because they only involve
characterization by three parameters which may be determined
theoretically, as shown here, or through experiment. Once determined,
these parameters may be used to construct \emph{any} two qubit gate.

\acknowledgments 
We would like to thank Gerard Milburn for support. CDH would like to
thank Mick Bremner, Jennifer Dodd, Henry Haselgrove and Tobias Osborne
for help and advice. HSG would like to acknowledge support from a
Hewlett-Packard Fellowship.  

\bibliography{kcGates}

\begin{thebibliography}{37}
\expandafter\ifx\csname natexlab\endcsname\relax\def\natexlab#1{#1}\fi
\expandafter\ifx\csname bibnamefont\endcsname\relax
  \def\bibnamefont#1{#1}\fi
\expandafter\ifx\csname bibfnamefont\endcsname\relax
  \def\bibfnamefont#1{#1}\fi
\expandafter\ifx\csname citenamefont\endcsname\relax
  \def\citenamefont#1{#1}\fi
\expandafter\ifx\csname url\endcsname\relax
  \def\url#1{\texttt{#1}}\fi
\expandafter\ifx\csname urlprefix\endcsname\relax\def\urlprefix{URL }\fi
\providecommand{\bibinfo}[2]{#2}
\providecommand{\eprint}[2][]{\url{#2}}

\bibitem[{\citenamefont{Grover}(1997)}]{Gro97}
\bibinfo{author}{\bibfnamefont{L.~K.} \bibnamefont{Grover}},
  \bibinfo{journal}{Phys. Rev. Lett.} \textbf{\bibinfo{volume}{79}},
  \bibinfo{pages}{325} (\bibinfo{year}{1997}).

\bibitem[{\citenamefont{Shor}(1997)}]{Sho97}
\bibinfo{author}{\bibfnamefont{P.~W.} \bibnamefont{Shor}},
  \bibinfo{journal}{SIAM Journal of Computing} \textbf{\bibinfo{volume}{26}},
  \bibinfo{pages}{1484} (\bibinfo{year}{1997}).

\bibitem[{\citenamefont{Knill et~al.}(2001)\citenamefont{Knill, Laflamme, and
  Milburn}}]{KLM01}
\bibinfo{author}{\bibfnamefont{E.}~\bibnamefont{Knill}},
  \bibinfo{author}{\bibfnamefont{R.}~\bibnamefont{Laflamme}}, \bibnamefont{and}
  \bibinfo{author}{\bibfnamefont{G.~J.} \bibnamefont{Milburn}},
  \bibinfo{journal}{Nature} \textbf{\bibinfo{volume}{409}}, \bibinfo{pages}{46}
  (\bibinfo{year}{2001}).

\bibitem[{\citenamefont{Gershenfeld and Chuang}(1997)}]{GC97}
\bibinfo{author}{\bibfnamefont{N.~A.} \bibnamefont{Gershenfeld}}
  \bibnamefont{and} \bibinfo{author}{\bibfnamefont{I.~L.}
  \bibnamefont{Chuang}}, \bibinfo{journal}{Science}
  \textbf{\bibinfo{volume}{275}}, \bibinfo{pages}{350} (\bibinfo{year}{1997}).

\bibitem[{\citenamefont{Cory et~al.}(1997)\citenamefont{Cory, Fahmy, and
  Havel}}]{CFH97}
\bibinfo{author}{\bibfnamefont{D.~G.} \bibnamefont{Cory}},
  \bibinfo{author}{\bibfnamefont{A.~F.} \bibnamefont{Fahmy}}, \bibnamefont{and}
  \bibinfo{author}{\bibfnamefont{T.~F.} \bibnamefont{Havel}},
  \bibinfo{journal}{Proc. Natl. Acad. Sci.} \textbf{\bibinfo{volume}{94}},
  \bibinfo{pages}{1634} (\bibinfo{year}{1997}).

\bibitem[{\citenamefont{Cirac and Zoller}(1995)}]{CZ95}
\bibinfo{author}{\bibfnamefont{J.~I.} \bibnamefont{Cirac}} \bibnamefont{and}
  \bibinfo{author}{\bibfnamefont{P.}~\bibnamefont{Zoller}},
  \bibinfo{journal}{Phys. Rev. Lett.} \textbf{\bibinfo{volume}{74}},
  \bibinfo{pages}{4091} (\bibinfo{year}{1995}).

\bibitem[{\citenamefont{Nakamura et~al.}(1998)\citenamefont{Nakamura, Pashkin,
  and Tsai}}]{NPT98}
\bibinfo{author}{\bibfnamefont{Y.}~\bibnamefont{Nakamura}},
  \bibinfo{author}{\bibfnamefont{Y.~A.} \bibnamefont{Pashkin}},
  \bibnamefont{and} \bibinfo{author}{\bibfnamefont{J.~S.} \bibnamefont{Tsai}},
  \bibinfo{journal}{Nature} \textbf{\bibinfo{volume}{398}},
  \bibinfo{pages}{786} (\bibinfo{year}{1998}).

\bibitem[{\citenamefont{Imamoglu et~al.}(1999)\citenamefont{Imamoglu,
  Awschalom, Burkard, DiVincenzo, Loss, Sherwin, and Small}}]{IAB+99}
\bibinfo{author}{\bibfnamefont{A.}~\bibnamefont{Imamoglu}},
  \bibinfo{author}{\bibfnamefont{D.~D.} \bibnamefont{Awschalom}},
  \bibinfo{author}{\bibfnamefont{G.}~\bibnamefont{Burkard}},
  \bibinfo{author}{\bibfnamefont{D.~P.} \bibnamefont{DiVincenzo}},
  \bibinfo{author}{\bibfnamefont{D.}~\bibnamefont{Loss}},
  \bibinfo{author}{\bibfnamefont{M.}~\bibnamefont{Sherwin}}, \bibnamefont{and}
  \bibinfo{author}{\bibfnamefont{A.}~\bibnamefont{Small}},
  \bibinfo{journal}{Phys. Rev. Lett.} \textbf{\bibinfo{volume}{83}},
  \bibinfo{pages}{4204} (\bibinfo{year}{1999}).

\bibitem[{\citenamefont{Kane}(1998)}]{Kan98}
\bibinfo{author}{\bibfnamefont{B.~E.} \bibnamefont{Kane}},
  \bibinfo{journal}{Nature} \textbf{\bibinfo{volume}{393}},
  \bibinfo{pages}{133} (\bibinfo{year}{1998}).

\bibitem[{\citenamefont{Honig}(1954)}]{Hon54}
\bibinfo{author}{\bibfnamefont{A.}~\bibnamefont{Honig}},
  \bibinfo{journal}{Phys. Rev.} \textbf{\bibinfo{volume}{96}},
  \bibinfo{pages}{254} (\bibinfo{year}{1954}).

\bibitem[{\citenamefont{Gordon and Bowers}(1958)}]{GB58}
\bibinfo{author}{\bibfnamefont{J.~P.} \bibnamefont{Gordon}} \bibnamefont{and}
  \bibinfo{author}{\bibfnamefont{K.~D.} \bibnamefont{Bowers}},
  \bibinfo{journal}{Phys. Rev. Lett.} \textbf{\bibinfo{volume}{1}},
  \bibinfo{pages}{10} (\bibinfo{year}{1958}).

\bibitem[{\citenamefont{Feher and Gere}(1959)}]{FG59}
\bibinfo{author}{\bibfnamefont{G.}~\bibnamefont{Feher}} \bibnamefont{and}
  \bibinfo{author}{\bibfnamefont{E.~A.} \bibnamefont{Gere}},
  \bibinfo{journal}{Phys. Rev.} \textbf{\bibinfo{volume}{114}},
  \bibinfo{pages}{1245} (\bibinfo{year}{1959}).

\bibitem[{\citenamefont{Feher}(1959)}]{Feh59}
\bibinfo{author}{\bibfnamefont{G.}~\bibnamefont{Feher}},
  \bibinfo{journal}{Phys. Rev.} \textbf{\bibinfo{volume}{114}},
  \bibinfo{pages}{1219} (\bibinfo{year}{1959}).

\bibitem[{\citenamefont{Honig and Stupp}(1960)}]{HSE60}
\bibinfo{author}{\bibfnamefont{A.}~\bibnamefont{Honig}} \bibnamefont{and}
  \bibinfo{author}{\bibfnamefont{E.}~\bibnamefont{Stupp}},
  \bibinfo{journal}{Phys. Rev.} \textbf{\bibinfo{volume}{117}},
  \bibinfo{pages}{69} (\bibinfo{year}{1960}).

\bibitem[{\citenamefont{Faulkner}(1969)}]{Fau69}
\bibinfo{author}{\bibfnamefont{R.~A.} \bibnamefont{Faulkner}},
  \bibinfo{journal}{Phys. Rev.} \textbf{\bibinfo{volume}{184}},
  \bibinfo{pages}{713} (\bibinfo{year}{1969}).

\bibitem[{\citenamefont{Chiba and Hirai}(1972)}]{CH72}
\bibinfo{author}{\bibfnamefont{M.}~\bibnamefont{Chiba}} \bibnamefont{and}
  \bibinfo{author}{\bibfnamefont{A.}~\bibnamefont{Hirai}}, \bibinfo{journal}{J.
  Phys. Soc. Japan} \textbf{\bibinfo{volume}{33}}, \bibinfo{pages}{730}
  (\bibinfo{year}{1972}).

\bibitem[{\citenamefont{Waugh and Slichter}(1988)}]{WJS88}
\bibinfo{author}{\bibfnamefont{J.~S.} \bibnamefont{Waugh}} \bibnamefont{and}
  \bibinfo{author}{\bibfnamefont{C.~P.} \bibnamefont{Slichter}},
  \bibinfo{journal}{Phys. Rev. B} \textbf{\bibinfo{volume}{37}},
  \bibinfo{pages}{4337} (\bibinfo{year}{1988}).

\bibitem[{\citenamefont{Goan and Milburn}(2000)}]{GM00}
\bibinfo{author}{\bibfnamefont{H.-S.} \bibnamefont{Goan}} \bibnamefont{and}
  \bibinfo{author}{\bibfnamefont{G.~J.} \bibnamefont{Milburn}},
  \bibinfo{journal}{Unpublished Manuscript}  (\bibinfo{year}{2000}).

\bibitem[{\citenamefont{Wellard}(2001)}]{Wel01}
\bibinfo{author}{\bibfnamefont{C.~J.} \bibnamefont{Wellard}},
  \bibinfo{journal}{PhD Thesis}  (\bibinfo{year}{2001}).

\bibitem[{\citenamefont{Wellard and Hollenberg}(2001)}]{WCJ01}
\bibinfo{author}{\bibfnamefont{C.~J.} \bibnamefont{Wellard}} \bibnamefont{and}
  \bibinfo{author}{\bibfnamefont{L.~C.~L.} \bibnamefont{Hollenberg}}
  (\bibinfo{year}{2001}), \eprint{quant-ph/0104055}.

\bibitem[{\citenamefont{Fowler et~al.}(2003)\citenamefont{Fowler, Wellard, and
  Hollenberg}}]{FWH03}
\bibinfo{author}{\bibfnamefont{A.~G.} \bibnamefont{Fowler}},
  \bibinfo{author}{\bibfnamefont{C.~J.} \bibnamefont{Wellard}},
  \bibnamefont{and} \bibinfo{author}{\bibfnamefont{L.~C.~L.}
  \bibnamefont{Hollenberg}}, \bibinfo{journal}{Phys. Rev. A}
  \textbf{\bibinfo{volume}{67}} (\bibinfo{year}{2003}).

\bibitem[{\citenamefont{Cummins and Jones}(2000)}]{CJ99}
\bibinfo{author}{\bibfnamefont{H.~K.} \bibnamefont{Cummins}} \bibnamefont{and}
  \bibinfo{author}{\bibfnamefont{J.~A.} \bibnamefont{Jones}},
  \bibinfo{journal}{New Journal of Physics} \textbf{\bibinfo{volume}{2}},
  \bibinfo{pages}{6} (\bibinfo{year}{2000}).

\bibitem[{\citenamefont{Tyco}(1983)}]{Tyc83}
\bibinfo{author}{\bibfnamefont{R.}~\bibnamefont{Tyco}}, \bibinfo{journal}{Phys.
  Rev. Lett.} \textbf{\bibinfo{volume}{51}}, \bibinfo{pages}{775}
  (\bibinfo{year}{1983}).

\bibitem[{\citenamefont{Wellard et~al.}(2002)\citenamefont{Wellard, Hollenberg,
  and Pauli}}]{WCJ01b}
\bibinfo{author}{\bibfnamefont{C.~J.} \bibnamefont{Wellard}},
  \bibinfo{author}{\bibfnamefont{L.~C.~L.} \bibnamefont{Hollenberg}},
  \bibnamefont{and} \bibinfo{author}{\bibfnamefont{H.~C.} \bibnamefont{Pauli}},
  \bibinfo{journal}{Phys. Rev. A} \textbf{\bibinfo{volume}{65}},
  \bibinfo{pages}{032303} (\bibinfo{year}{2002}).

\bibitem[{\citenamefont{Kraus and Cirac}(2001)}]{KC01}
\bibinfo{author}{\bibfnamefont{B.}~\bibnamefont{Kraus}} \bibnamefont{and}
  \bibinfo{author}{\bibfnamefont{J.~I.} \bibnamefont{Cirac}},
  \bibinfo{journal}{Phys. Rev. A} \textbf{\bibinfo{volume}{63}},
  \bibinfo{pages}{062309} (\bibinfo{year}{2001}).

\bibitem[{\citenamefont{Hammerer et~al.}(2002)\citenamefont{Hammerer, Vidal,
  and Cirac}}]{HVC02}
\bibinfo{author}{\bibfnamefont{K.}~\bibnamefont{Hammerer}},
  \bibinfo{author}{\bibfnamefont{G.}~\bibnamefont{Vidal}}, \bibnamefont{and}
  \bibinfo{author}{\bibfnamefont{J.~I.} \bibnamefont{Cirac}},
  \bibinfo{journal}{Phys. Rev. Lett.} \textbf{\bibinfo{volume}{88}},
  \bibinfo{pages}{237902} (\bibinfo{year}{2002}), \eprint{quant-ph/0205100}.

\bibitem[{\citenamefont{Khaneja et~al.}(2001)\citenamefont{Khaneja, Brockett,
  and Glaser}}]{KBG00}
\bibinfo{author}{\bibfnamefont{N.}~\bibnamefont{Khaneja}},
  \bibinfo{author}{\bibfnamefont{R.}~\bibnamefont{Brockett}}, \bibnamefont{and}
  \bibinfo{author}{\bibfnamefont{S.}~\bibnamefont{Glaser}},
  \bibinfo{journal}{Phys. Rev. A} \textbf{\bibinfo{volume}{63}},
  \bibinfo{pages}{032308} (\bibinfo{year}{2001}).

\bibitem[{\citenamefont{Bremner et~al.}(2002)\citenamefont{Bremner, Dawson,
  Dodd, Gilchrist, Harrow, Mortimer, Nielsen, and Osborne}}]{BDD+02}
\bibinfo{author}{\bibfnamefont{M.~J.} \bibnamefont{Bremner}},
  \bibinfo{author}{\bibfnamefont{C.~M.} \bibnamefont{Dawson}},
  \bibinfo{author}{\bibfnamefont{J.~L.} \bibnamefont{Dodd}},
  \bibinfo{author}{\bibfnamefont{A.}~\bibnamefont{Gilchrist}},
  \bibinfo{author}{\bibfnamefont{A.~W.} \bibnamefont{Harrow}},
  \bibinfo{author}{\bibfnamefont{D.}~\bibnamefont{Mortimer}},
  \bibinfo{author}{\bibfnamefont{M.~A.} \bibnamefont{Nielsen}},
  \bibnamefont{and} \bibinfo{author}{\bibfnamefont{T.~J.}
  \bibnamefont{Osborne}}, \bibinfo{journal}{Phys. Rev. Lett.}
  \textbf{\bibinfo{volume}{89}}, \bibinfo{pages}{247902}
  (\bibinfo{year}{2002}).

\bibitem[{\citenamefont{Kettle et~al.}(2003)\citenamefont{Kettle, Goan, Smith,
  Hollenberg, Pakes, and Wellard}}]{KG03}
\bibinfo{author}{\bibfnamefont{L.}~\bibnamefont{Kettle}},
  \bibinfo{author}{\bibfnamefont{H.-S.} \bibnamefont{Goan}},
  \bibinfo{author}{\bibfnamefont{S.~C.} \bibnamefont{Smith}},
  \bibinfo{author}{\bibfnamefont{L.~C.~L.} \bibnamefont{Hollenberg}},
  \bibinfo{author}{\bibfnamefont{C.~I.} \bibnamefont{Pakes}}, \bibnamefont{and}
  \bibinfo{author}{\bibfnamefont{C.}~\bibnamefont{Wellard}},
  \bibinfo{journal}{Paper in preparation}  (\bibinfo{year}{2003}).

\bibitem[{\citenamefont{Becker}(2000)}]{Bec00}
\bibinfo{author}{\bibfnamefont{E.~D.} \bibnamefont{Becker}},
  \emph{\bibinfo{title}{High {R}esolution {NMR}}} (\bibinfo{publisher}{Academic
  Press}, \bibinfo{address}{San Diego}, \bibinfo{year}{2000}),
  \bibinfo{edition}{3rd} ed.

\bibitem[{\citenamefont{Slichter}(1990)}]{Sli78}
\bibinfo{author}{\bibfnamefont{C.~P.} \bibnamefont{Slichter}},
  \emph{\bibinfo{title}{Principles of {M}agnetic {R}esonance}}
  (\bibinfo{publisher}{Springer-Verlag}, \bibinfo{address}{Berlin},
  \bibinfo{year}{1990}), \bibinfo{edition}{3rd} ed.

\bibitem[{\citenamefont{Dodd et~al.}(2002)\citenamefont{Dodd, Nielsen, Bremner,
  and Thew}}]{DNB+01}
\bibinfo{author}{\bibfnamefont{J.~L.} \bibnamefont{Dodd}},
  \bibinfo{author}{\bibfnamefont{M.~A.} \bibnamefont{Nielsen}},
  \bibinfo{author}{\bibfnamefont{M.~J.} \bibnamefont{Bremner}},
  \bibnamefont{and} \bibinfo{author}{\bibfnamefont{R.~T.} \bibnamefont{Thew}},
  \bibinfo{journal}{Phys. Rev. A} \textbf{\bibinfo{volume}{65}},
  \bibinfo{pages}{040301} (\bibinfo{year}{2002}).

\bibitem[{\citenamefont{Haselgrove}(2002)}]{Has02}
\bibinfo{author}{\bibfnamefont{H.}~\bibnamefont{Haselgrove}},
  \bibinfo{journal}{Private Communication}  (\bibinfo{year}{2002}).

\bibitem[{\citenamefont{DiVincenzo}(1995)}]{DiV95}
\bibinfo{author}{\bibfnamefont{D.}~\bibnamefont{DiVincenzo}},
  \bibinfo{journal}{Phys. Rev. A} \textbf{\bibinfo{volume}{51}},
  \bibinfo{pages}{1015} (\bibinfo{year}{1995}).

\bibitem[{\citenamefont{Shor}(1996)}]{Sho96}
\bibinfo{author}{\bibfnamefont{P.~W.} \bibnamefont{Shor}},
  \bibinfo{journal}{37th Annual Symposium on Fundamentals of Computer Science,
  Proceedings of} pp. \bibinfo{pages}{56--65} (\bibinfo{year}{1996}).

\bibitem[{\citenamefont{Nielsen and Chuang}(2001)}]{NC01}
\bibinfo{author}{\bibfnamefont{M.~A.} \bibnamefont{Nielsen}} \bibnamefont{and}
  \bibinfo{author}{\bibfnamefont{I.~L.} \bibnamefont{Chuang}},
  \emph{\bibinfo{title}{Quantum {C}omputation and {Q}uantum {I}nformation}}
  (\bibinfo{publisher}{Cambridge University Press},
  \bibinfo{address}{Cambridge}, \bibinfo{year}{2001}), \bibinfo{edition}{2nd}
  ed.

\bibitem[{\citenamefont{Loss and DiVincenzo}(1998)}]{LD98}
\bibinfo{author}{\bibfnamefont{D.}~\bibnamefont{Loss}} \bibnamefont{and}
  \bibinfo{author}{\bibfnamefont{D.~P.} \bibnamefont{DiVincenzo}},
  \bibinfo{journal}{Phys. Rev. A} \textbf{\bibinfo{volume}{57}},
  \bibinfo{pages}{120} (\bibinfo{year}{1998}).

\end{thebibliography}

\end{document}